**Attenuation of marine seismic interference noise employing a customized U-Net**

Jing Sun[1,2*], Sigmund Slang[1,2], Thomas Elboth[2], Thomas Larsen Greiner[1,3], Steven McDonald[2], Leiv-J Gelius[1]

[1] University of Oslo, Department of Geosciences, Sem Sælands vei 1, 0371 Oslo, Norway

[2] CGG

[3] Lundin Norway AS, Strandveien 4, 1366 Lysaker, Norway

**ABSTRACT**

Marine seismic interference noise occurs when energy from nearby marine seismic source vessels is recorded during a seismic survey. Such noise tends to be well preserved over large distances and cause coherent artifacts in the recorded data. Over the years, the industry has developed various denoising techniques for seismic interference removal, but although well performing they are still time-consuming in use. Machine-learning based processing represents an alternative approach, which may significantly improve the computational efficiency. In case of conventional images, autoencoders are frequently employed for denoising purposes. However, due to the special characteristics of seismic data as well as the noise, autoencoders failed in the case of marine seismic interference noise. We therefore propose the use of a customized U-Net design with element-wise summation as part of the skip-connection blocks to handle the vanishing gradient problem and to ensure information fusion between high- and low-level features. To secure a realistic study, only seismic field data were employed, including 25000 training examples. The customized U-Net was found to perform well leaving only minor residuals, except for the case when seismic interference noise comes from the side. We further demonstrate that such noise can be treated by slightly increasing the depth of our network. Although our customized U-Net does not outperform a standard commercial algorithm in quality, it can (after proper training) read and process one single shot gather in approximately 0.02s. This is significantly faster than any existing industry denoising algorithm. In addition, the proposed network processes shot gathers in a sequential order, which is an advantage compared with industry algorithms that typically require a multi-shot input to break the coherency of the noise.

**INTRODUCTION**

Machine learning (ML) has recently received a great deal of attention within the geoscience community following its successful application within fields such as image recognition and natural language processing. Easier access to affordable and powerful hardware (CPU and GPU) solutions together with user-friendly open-source software such as TensorFlow (Abadi *et al.* 2016) and Pytorch (Paszke *et al.* 2017), has been the key to the accelerated use of ML within various areas of technology. In this paper, we investigate the feasibility of using ML-based techniques to attenuate marine seismic interference (SI) noise. Marine SI noise is a type of coherent noise which occurs when unwanted acoustic energy from nearby marine seismic source vessels not linked to the survey is recorded. It is very common, especially in areas such as the North Sea, where the exploration activities peak during summer as weather conditions are often unsuitably during other periods (Elboth, Presterud and Hermansen 2010). SI noise can typically be observed as strong coherent events in data recordings, and tend to be well preserved over large distances (Jansen, Elboth and Sanchis 2013).

One way to minimize the influence of SI noise during a survey is to initiate timesharing. This means that only one vessel shoots at a time while the other vessels are on standby. Such timesharing has been common for contractor companies in the Gulf of Mexico (Akbulut *et al.* 1984) and in the North Sea (Laurain, Ruiz-Lopez and Eidsvig 2014). However, this type of acquisition is both costly and inefficient since it often results in substantial downtime and cost overrun. The more attractive and cost-effective alternative is therefore to continue acquiring data with SI noise included and then attenuate it later in processing.

Although marine SI noise is coherent in the shot domain, it is often more incoherent in the common-receiver or common-offset domain. It can therefore be attenuated by the use of f-x prediction filters after resorting the data to one of these domains (Wang *et al.* 1989). Gulunay and Pattberg (2001a, b) proposed threshold-guided detection of noisy shots in the frequency domain followed by the use of f-x prediction filters to remove inline SI noise. Guo and Lin (2003) investigated the similar use of inline prediction filters but replaced straight subtraction with an adaptive subtraction process. Kommedal, Semb and Manning (2007) discussed the attenuation of SI noise present in multicomponent 4D data for permanently installed arrays. Gulunay (2008) proposed an approach to identify and attenuate SI noise through the use of propagation models. Jansen *et al.* (2013) introduced similar concepts based on an automatic vector field estimation of the local moveout of the SI noise. A more recent industry technique to remove marine SI noise, has been the use of $\tau-p$ transformation followed by random noise attenuation on common-slowness gathers (Elboth *et al.* 2010). Using this approach, Laurain *et al.* (2014) demonstrated that SI noise only poses a real problem if it arrives broadside where it has similar moveout to the useful seismic reflection data and therefore data acquisition in this situation should be avoided in the field.

Deep learning is a class of machine learning algorithms that use multiple layers to progressively extract higher-level features from raw input (Deng and Yu 2014). Most modern deep learning models are based on artificial neural networks (ANNs). In deep learning, each level learns to transform its input data into an increasingly abstract and composite representation. A convolutional neural network (CNN) is a class of ANNs, which was inspired by biological processes in that the connectivity pattern between neurons resembles the organization of the visual cortex (Fukushima 2007; Hubel and Wiesel 1968; Matsugu *et al.* 2003). During the training process of a CNN, a hierarchy of local filters learn to extract the essential features of the training data relevant to the application in question. In recent years, more and more attempts have been made to employ CNN in various fields of science and engineering, including geophysics.

The first successful applications of CNNs in geophysics were within seismic image classification and interpretation. Rentsch *et al.* (2014) trained a neural network to identify SI noise for towed-streamer acquisitions. Waldeland *et al.* (2018) demonstrated how CNNs could be used to classify different seismic textures with special emphasis on salt bodies. Qian *et al.* (2018) applied a deep convolutional autoencoder (DCAE) network in seismic facies recognition based on prestack seismic data. Xiong *et al.* (2018) trained a CNN to automatically detect and map fault zones using 3D seismic images, whereas Wu *et al.* (2019) proposed the use of CNNs to pick the first arrivals of microseismic events. Baardman (2018) and Baardman and Tsingas (2019) proposed the use of a CNN to classify data patches in a "blended" and "non-blended" class. A second, regression based, CNN was then employed to deblend the "blended" patches. However, only synthetic data were considered.

Within the area of seismic data interpolation and reconstruction, Wang *et al.* (2018a, b) introduced an 8-layer residual learning network (ResNet) based on CNNs to interpolate seismic data without aliasing. Mandelli *et al.* (2018, 2019) proposed to reconstruct missing seismic traces in the prestack domain by employing a convolutional autoencoder (AE). They applied this network to solve the joint problem of synthetic data interpolation and Gaussian-noise attenuation. Although not yet fully investigated, the use of CNNs in the field of seismic noise suppression is also underway. Liu *et al.* (2018) managed to remove random noise from a 3D poststack seismic dataset using a 3D CNN architecture. Ma (2018) used a CNN to attenuate multiples, linear noise and random noise simultaneously on a controlled dataset where both the training and test datasets were computed from the same model. Slang *et al.* (2019) employed marine seismic field data and demonstrated successful applications within deblending and denoising using CNNs.

In this paper, the feasibility of employing CNNs within the area of SI-denoising is investigated. To ensure that this study is as realistic as possible, field data are employed. The data diversity is properly addressed by using 25000, 3000 and 1500 images for training, validation and testing respectively.

The paper is organized as follows. Firstly, the characteristics and corresponding denoising challenges of different types of SI noise are presented. This basic introduction is followed by a discussion of differences between conventional images and seismic data using the Autoencoder (AE) as an example. In the second section, the basic concepts of CNNs are introduced, followed by a section describing and discussing the customized U-Net architecture used in this study. Examples of employing the proposed network to attenuate different types of SI noise are presented in the next section. We further demonstrate that also broadside SI noise can be treated by slightly increasing the depth of our network (see Appendix A). In the section to follow, we compare the denoised seismic stacks obtained using respectively the proposed network and a standard industry SI-denoising algorithm. Since the network was trained on data from a different geological area, this comparison study also serves to illustrate the robustness of the proposed approach. Finally, a discussion and a set of conclusions are presented.

**CHARACTERISTICS OF SEISMIC INTERFERENCE NOISE**

SI noise mainly travels in the water column with the seafloor and sea surface acting as reflectors channeling the energy. A small portion of the energy may propagate in subsurface layers, but will make little to no impact in the recorded data due to high attenuation factors. In relative shallow waters, for example in the North Sea, SI noise can often be observed from sources 100 to 120km away. However, in deeper water environments, for example the Gulf of Mexico, SI is normally not observed from sources beyond 60 to 80km. Compared to seismic random noise and Gaussian noise, SI noise is a significant problem for seismic contractor companies, causing artefacts in the data. These artefacts have a characteristic appearance as coherent linear (and possibly non-linear) events with high amplitude (Akbulut *et al.* 1984). The noise is typically clearly visible as it appears at different arrival times in each record. The angles of incidence for SI noise may differ from survey to survey depending on the relative placement of its origin to the receiver. Because of this, SI noise might overlap with sub-surface layer reflections which can have significantly lower amplitudes. SI can, therefore be harmful to a number of processing operations such as deghosting, demultiple, velocity estimations and amplitude versus offset (AVO) analysis (Gulunay, Magesan and Baldock 2004).

Figure 1 illustrates SI noise contaminated shot gathers with varying angle and distance to the source vessel. Figures 1a and 1b show SI noise coming from ahead of the recording vessel. The alignment of the noise is similar to the seismic signal since they originate from the same direction. Compared to Fig. 1b, the SI-noise in Fig. 1a is almost linear due to the much longer distance between the source and receiver vessels. The further away the source, the more linear the SI noise appears in the shot domain. It is hard to give a specific distance as this is a gradual shift. In general, the strictly linear structure of SI noise only yields when the origin of noise is further away than approximately 20km to 40km. For SI noise events originating from sources closer than approximately 20km, the events appear more curved.

Figure 1c shows SI coming from the side of the receiver vessel within close proximity (6km). The noise moveout has a significant curvature in this case. The amplitude of the noise is high and can mask the underlying geological reflection data. This type of SI noise appears kinematically similar to reflection data and is therefore difficult to remove. It is the most harmful SI noise and should be avoided as much as possible when acquiring data. Figure 1d shows SI coming from astern with a large distance to the source vessel. The noise appears similar to Fig. 1a, but mirrored along the offset direction due to distance of origin. SI noise traveling in the water column may have slightly changing velocity depending on level of salinity and temperature in the water, but it can be approximated to 1500m/s.

**Differences between seismic data and conventional images**

When applying ANNs to a specific problem, there are multiple factors to take into account. If certain prerequisites are not met, the model might under-perform, overfit or behave in other sub-optimal ways. One of the most important factors when designing a model is to fit the model to the desired task and dataset. Marine seismic data are represented as matrices, but can also be regarded as images. It is therefore likely to adapt a model in a similar way as one might when doing image processing with machine learning. There are, however significant differences between conventional images and seismic data, as listed in Table 1.

To further illustrate these differences, we made use of the Autoencoder (AE) network structure commonly used for image processing in machine learning. AEs have proven to yield good results when used to denoise conventional images (Gondara 2016; Xie, Xu and Chen 2012).

The fundamental concept behind an AE is to represent data at a latent level. To do this, the network has to compress the data, thus separate important features, while omitting less important features. After the data are downscaled to the latent level, the network tries to recreate the data by minimizing the difference between the ground truth and the output of the AE. The characteristics of an AE fit well with noise removal. If the noise appearance varies from image to image, it will be regarded as an unimportant feature which can therefore be removed. Due to above reason, we decided to test the capability of an AE to attenuate SI noise. The AE network design used in this paper is visualized in Fig. 2.

When training a network, ground truth data should be available. Thus, we artificially constructed a SI-noise contaminated marine seismic data set by randomly blending 482 SI noise-free shot gathers from the North Sea and 800 records containing almost pure SI noise recorded from different directions. However, various types of noise always contaminate marine seismic data acquired in the field. Thus, in general, the use of advanced processing techniques is needed in order to establish the ground truth. When combined, the training, validation and test data sets consisted of 25000, 3000 and 1500 images respectively.

Seismic datasets with a 2ms sampling rate hold 500 time samples for each second. The original data have a record time of 9 seconds two-way traveltime (TWT) resulting in 4500 time samples for each trace. This is considered large to the network and will cause computational slowness. Thus, we down-sampled the data to 4ms and cut it down to 6 seconds to further reduce computation time and memory consumption. The final input size of each seismic image is then 1500×256, as far offset traces also were omitted to increase computational efficiency.

We trained the AE network shown in Fig. 2 for 40 epochs until the loss converged towards a stable value. The denoising results together with the corresponding frequency spectra are displayed in Fig. 3. In the same figure, also the training and validation losses are shown. The actual output from the network, Fig. 3c, shows that some SI noise have been attenuated but with much of the original noise still left. The zoomed section around 5 seconds TWT, box iii, highlights an area with significant noise residual. The noise overshadows the underlying signal and it is difficult to see whether the signal is kept intact due to the low signal-to-noise ratio (SNR). Box vi in Fig. 3d highlights the same part in the difference between the ground truth and the output showing weak residual geological signals and high amounts of residual noise. The residual noise is pixelated which is likely an artefact introduced by the compression in the model. Boxes i and iv in Fig. 3c and Fig. 3d respectively highlight an area around 2 to 3 seconds TWT with strong reflections and multiples. Box i shows a seemingly clear and noise free reconstruction of this section, but on comparison with box iv in the difference shows that there is substantial signal loss in this area. The box ii highlighted in Fig. 3c around 4 seconds TWT shows a high SNR as there are little noise present in this area. However, direct comparison with box v in Fig. 3d shows that some geological information has been lost. Fig. 3e shows the frequency spectrum of the ground truth and output. It is clear that the AE struggles with recreating the frequency band as it performs increasingly worse with increasing frequency from around 25Hz.

Thus, direct use of a typical machine learning design tailored for denoising of conventional images do not necessarily work in an optimal sense if employed to seismic data. The reasons are most likely due to a combination of several issues. Among them, the differences in main characteristics between seismic data and a conventional image as already discussed briefly, but also the character of the noise itself. In many applications reported in the literature, the type of noise present in conventional images is of a Gaussian or random character. In such cases, the signal and the noise easily separate during a compression phase as part of an AE. However, in case of SI noise, being coherent, the compression phase may harm the signal bandwidth due to overlap in character between signal and noise space. We also mention that the dataset we used had SI noise from many different directions, arriving at many different times. It may be that even though we used 25000 training images, this was still not sufficient for the network to learn to optimally represent seismic data at the latent (compressed) level in the AE.

**BASIC CONCEPT OF A CONVOLUTIONAL NEURAL NETWORK**

The definition of a convolutional neural network (CNN) is rather loose. According to Goodfellow, Bengio and Courville (2016) a CNN has to have at least one convolutional layer.

Each network has to be tailor made for each specific application. However, there are certain guidelines that can be followed. As briefly discussed in the previous section, the color depth of a seismic image is one, meaning that marine seismic data can be represented as 2D matrices, different from the 3D representation of colored images. Before the different building blocks of a neural network can be explained in more detail, certain basic definitions need to be reviewed.

**Convolutional layer**

A convolutional layer is a layer which, suggested by its name, convolves the input. Each convolutional layer has assigned a set of filters (kernels) where all have the same size. These filters are normally square matrices consisting of decimal values with a chosen size. Let $A \in \mathbb{R}^{3 \times 3}$ denote the input image with elements $a_{k,l}$ for $k,l \in \{0,2\}$, $O \in \mathbb{R}^{2 \times 2}$ denote the output image with elements $O_{m,n}$ for $m,n \in \{0,1\}$ and $W \in \mathbb{R}^{2 \times 2}$ denote the filter kernel with weights $w_{i,j}$ for $i,j \in \{0,1\}$. Fig. 4 gives an example of how the 2×2 filter works on the 3×3 image with stride equal to 1, to give the output image (with mirrored kernel to ensure convolution). The stride is defined by the distance between two consecutive positions of the filter kernel (Dumoulin and Visin 2018). The 2D convolution operation (Fig. 4a) can be represented by the neural network configuration (Fig. 4b) where the filter weights are represented by color coding. We can take the orange square inside the red box as an example. In this case, the result from the convolution comes from the linear combination $w_{1,1}a_{1,1} + w_{1,0}a_{1,2} + w_{0,1}a_{2,1} + w_{0,0}a_{2,2} = o_{1,1}$, where the kernel is linked to the orange neuron inside the red circle in the network by the gray arrow. As shown in Fig. 4b, only four blue neurons in the previous layer are connected through weights (filter coefficients) with the orange neuron inside the red circle (with weights being color coded according to the color scheme chosen for the filter coefficients).

A feedforward neural network consists of basic units represented by the neurons that are stacked into layers, with the output of one layer serving as the input for the next one. The complete neural network can be viewed as a complicated nonlinear transformation of the input into a predicted output that depends on the learnable weights and biases of all the neurons in the input layer, the hidden layers and output layer (Mehta *et al.* 2019).

Consider a training data set $\{T_i, \tilde{T}_i\}_{i=1}^{M}$ where $T$ and $\tilde{T}$ defines the clean (ground truth) and SI contaminated data respectively. We construct and train a function (network) $f_{W,b}: \tilde{T}_i \to T_i$ which in our case is based on a conventional feed forward CNN architecture. Let $N_l$ denote the number of features and $k_l = 1, 2, ..., N_l$ denote the $k$'th convolutional filter in layer $l = 1, 2, ..., L$. The feature mapping from one arbitrary layer to the next can be summarized by the expression

$$Z_{k_l}^{[l]} = \sum_{k_{l-1}=1}^{N_{l-1}} \left( W_{k_l}^{[l]} * A_{k_{l-1}}^{[l-1]} \right) + B_{k_l}^{[l]}, \tag{1}$$

where $W_{k_l}^{[l]} \in \sim^{f_{l-1} \times f_{l-1}}$ contains the weights and $B_{k_l}^{[l]}$ is a matrix of same size as $Z_{k_l}^{[l]}$ containing the biases. The notation $*$ denotes the convolutional process.

The convolution process of going from an arbitrary layer $l-1$ to layer $l$ represented by the first term on the right-hand-side of equation (1) is illustrated in Fig. 5 for one single element in the matrix $Z_{k_l}^{[l]}$. The matrix $W_{k_l}^{[l]}$ represents the $k_l$'th filter kernel spanning all activations in layer $l-1$ producing each of the $k_l = 1, 2, ..., N_l$ feature maps in layer *l*.

**Activation function**

In equation (1), $A_{k_{l-1}}^{[l-1]}$ represents one of the activations from the previous layer. The activation in ML is defined by a non-linear transformation on all $k_l = 1,2,...,N_l$ mappings of the feature maps, and defines the output from layer $l-1$ to layer $l$. The activation of layer $l$ can be represented by the general expression

$$A_{k_l}^{[l]} = \varphi^{[l]}(Z_{k_l}^{[l]}), \tag{2}$$

where $\varphi^{[l]}$ is the non-linear function. In our case, we chose the Leaky Rectified Linear Unit or Leaky ReLU (Maas, Hannun and Ng 2013). It is a commonly used version of ReLU which adjusts for the vanishing gradient problem. It introduces a small slope instead of forcing $\varphi^{[l]} = 0$ when $Z_{k_l}^{[l]} < 0$ and is defined as

$$\varphi^{[l]} = \max(Z_{k_l}^{[l]}, \alpha Z_{k_l}^{[l]}) \tag{3}$$

where $\alpha$ typically is set to 0.01 in conventional image processing cases. In the seismic case, however, we observed that the slope value $\alpha$ seems to benefit from being larger. In this study we therefore employed a value of 0.3. Leaky ReLU has the same characteristics as ReLU, but a reduced chance of causing "Dead Neurons" as the gradient is non-zero for $Z_{k_l}^{[l]} < 0$ (Maas *et al.* 2013). In the case where input to a ReLU with its weights is negative, the output will be 0, causing the gradient also to be 0. If instead Leaky ReLU is used, the gradient will never be 0 and this problem is avoided.

**Loss function**

Loss function is a key part of the network estimating how well the model is performing. Each time a batch (a term used in ML which means a group of training samples) is passed through the network, the loss is calculated and the parameters are updated accordingly. If the loss function is sub-optimal, the model might break down, fluctuate or take more time than strictly needed before it converges. In this study we tested several loss functions and found that minimum absolute error (MAE) gave the overall best results.

The predicted denoised image output from the network can be expressed as

$$\hat{T} = \varphi \left[ \sum_{k_{L-1}=1}^{N_{L-1}} \left( W_{k_L}^{[L]} * A_{k_{L-1}}^{[L-1]} \right) + B_{k_L}^{[L]} \right]. \tag{4}$$

The corresponding MAE loss function can be expressed by a L1-norm

$$L_1(W, B) = \left\| T - \hat{T} \right\|_1. \tag{5}$$

where $\hat{T}$ is the predicted 'noise-free' image and $T$ is the ground truth. MAE is not sensitive to outliers and will not penalize high errors. This means that MAE is more robust than mean square error (MSE) (Willmott and Matsuura 2005). There is, however, a downside when using MAE: its gradient has exactly the same slope as long as the difference is not equal to 0. This might cause issues during training where the network might struggle with converging to a minimum due to high gradients.

**Optimizer**

The optimizer is also an important building stone in a neural network. It is used alongside backward propagation and changes the weights and biases based on the gradient of the loss function. In essence, an optimizer has two tasks: finding the direction in which to move the weights and biases and the distance in which to move them (Mitchell 1997). There are multiple different optimizers, but most have one thing in common: they are based on gradient descent. In order to find the weights and biases that minimize equation (5), we train the network using a first-order gradient method for stochastic optimization, known as Root Mean Square propagation or RMSprop (Tieleman and Hinton 2012).

**Pooling layer**

A pooling layer is a layer constructed from local information with similarity to convolutional layers. The difference is that pooling layers down-sample the volume, while a convolutional layer does operations filter-wise. Pooling layers can be split into three different categories: max, average and min pooling, see Fig. 6. We assume the pooling has a filter size of 2×2. For each region, the layer does a specific operation based on the type of pooling. The pooling types will map 4 values to one output value where: max pooling sets the output value to be the largest of the 4 values, min pooling sets the output value to be the smallest of the 4 values and average pooling computes the average of the 4 values. This process can be viewed in Fig. 6 where different regions are color coded to ease the understanding. The output of a 2×2 pooling with stride 2 is a down-sampling by a factor of 2 in each direction.

Pooling is much used in machine learning, and the most common type is max pooling because large values tend to be most important when using images. Max pooling is a way for the network to extract the sharpest features of an image, thus the subsequent layers after a max pooling layer will have an image consisting mainly of important features and less abundant information. Reducing the size of the input volume is also a way to increase computational speed. The downside to using pooling is that whenever down-sampling is implemented, information is lost. For networks where images are down-sampled and up-sampled before output, the end results tend to be blurry. Max pooling is therefore a good way for the network to learn sharp features, but might cause lacking output due to information loss.

**Upsampling layer**

Upsampling layer, often called deconvolution, is essentially a transposed/inverse convolution. The layer operates in a similar manner to convolutional layer, but instead of scaling an image down, the image is scaled up. The output pixel in a normal convolution is the summation of the elementwise multiplication between input region and filter values. The transposed convolution works differently. It multiplies a value in the input with the filter values and copies it to multiple pixels, Fig. 7. The output from an upsampling layer will be weighted copies of the filter, weighted by the input. If upsampling by a factor of 2, the filter will move 2 pixels in the output for every corresponding pixel in the input. Overlapping regions in the output will be summed (Li, Johnson and Young 2017).

**Skip connection**

A skip connection represents a connection between an early layer and a later layer in a network, thus jumping over all layers in between. Such a connection can be formed employing either concatenation or element-wise summation (cf. Fig.8). The frequently used U-Net (Ronneberger, Fischer and Brox 2015) utilizes skip connections primarily through concatenation. ResNet (Residual Network) (He *et al.* 2016) is another well-known network architecture employing skip connections. However, in case of ResNet element-wise summation is employed. In the customized U-Net proposed in this paper, we take inspiration from ResNet and introduce a summation approach to skip connection instead of the concatenation operation employed in a standard U-Net. In this way, the vanishing gradient problem is avoided since the gradient flow is uninterrupted. In addition, skip connection ensures information fusion between high- and low-level features. This is beneficial in case of a network structure with contracting paths like our customized U-Net where the unwanted information loss during downscaling can be counterbalanced by the skip connections.

Figures 8a and 8b visualize the mathematical operation of using respectively element-wise summation and concatenation to connect a general layer $l-3$ and a deeper layer $l$.

Moreover, $Z^{[l-3]}$ and $Z^{[l]}$ represent the collections of feature maps from layer $l-3$ and layer $l$, respectively. Element-wise summation requires that the size and number of feature maps are the same for the two layers. This is different from concatenation where all feature maps are collected without any summation involved.

**A CUSTOMIZED U-NET FOR ATTENUATION OF SEISMIC INTERFERENCE NOISE**

It has already been demonstrated that the use of a conventional Autoencoder (AE) did not perform well when applied to the problem of SI-noise removal from marine seismic data. This is because spatial information from the input images will be gradually lost in the encoding process due to the overlap in characteristics of coherent noise and signal. To minimize possible loss in the signal part due to downscaling (compression), two other network architectures were investigated. The first architecture studied was different realizations of a No Downscaling CNN (NDCNN). In a NDCNN, downscaling is avoided so the seismic image is full-size throughout the network to reduce potential blurring and precision loss. This results in a large model requiring significant computational power and about 12GB GPU memory to train on a single image. The overall training process took several days on a modern GPU (6GB×2). However, we note that once the network is trained, the actual denoising of a single shot gather is done in less than a second. The results obtained using a NDCNN were quite encouraging, but the network performed less optimal in cases with a low SNR (Slang *et al.* 2019). In order to reduce the training time and possibly improve the denoising results further, a second class of CNN network architectures was investigated, namely the U-Net (Ronneberger *et al.* 2015). This network structure was originally proposed for biomedical image segmentation.

The U-Net architecture has the encoding-decoding process in common with the autoencoder. It consists of a contracting path to capture context and a more or less symmetric expanding path that enables precise localization (Ronneberger *et al.* 2015). A major advantage of the U-Net is the use of skip connections to compensate for the loss of information during the contracting path. We propose a slight modification of the standard U-Net by replacing concatenation with element-wise summation at every skip connection point. This avoids doubling the number of feature maps as well as the issue of vanishing gradients.

Figure 9 shows an example of using element-wise summation of information from a shallow layer with that from a deeper layer to ensure information fusion between high- and low-level features. Especially when a feature map from a deeper layer contains only weak information such as in Fig. 9e, the element-wise summation seems to enhance the contrast between the useful signal and noise.

Our customized U-Net combines the characteristics of both NDCNN and AE. It compresses images to a latent level in the same manner as an AE network, thus removing random features in the data. The data are then resampled back to original size where less noise is present. The compression is a good way of removing noise, but it might cause unwanted loss. NDCNN was a way of counteracting the potential loss from AE, but a model with no downscaling will demand more processing power. The customized U-Net utilizes the characteristic downscaling as seen in an AE, but with the ability to pass on higher resolution features between each upscaling by the use of skip connections.

Figure 10 visualizes the architecture of the customized U-Net used in this paper. The network consists of 7 convolutional layers each followed by a Leaky ReLU, except for the last layer which has no activation function. The first two convolutional layers employ max pooling to reduce the size of the data. There are two consecutive layers at latent level, where the last one employs upsampling. Before passed to the next layer, element-wise summation is employed combining data from the previous layer of same size. This is repeated for the next upscaling process, as seen in Fig. 10. Once the data reach original size, they are passed through a convolutional layer, and then fed to the output layer. The first two convolutional layers employ large filters to capture large scale features.

Training the NDCNN type of network required more than 100 hours to run 40 epochs on a fairly modern GPU, and the results still had much residuals left. This customized U-Net structure contained the compression structure of an AE, but had element-wise summation inheriting from less deep higher-resolution layers to counter for unwanted information loss during compression. The new network design required only 7.5 hours for 40 epochs on the same GPU. The processing GPU cluster consisted of eight nodes where 4 of the nodes were assigned to this work. Each node has the following specifications:

– CPU: Intel Xeon CPU E5-160 v3, 3.50 GHz, 10Mb cache, (8x)

– GPU: Nvidia GeForce GTX 1080, 8GB, 1.73 GHz, (2x)

– HDD: 62 GB storage (per user)

– OS: Debian 7, Wheezy

Having introduced the key components of the customized U-Net, we give a more detailed discussion of the various main design choices.

**Choice of activation functions**

Activation functions are introduced to the network to create non-linearity and thus solving problems that are governed by a higher-order equation. The four different activation functions tested were: TanH, ReLU, Leaky ReLU with $\alpha = 0.01$ and Leaky ReLU with $\alpha = 0.3$. TanH gave poorest result but also ReLU failed creating patterns of blank zones as shown in Fig. 11b. The performance of Leaky ReLU with a standard parameter choice of $\alpha = 0.01$ was slightly better, but the same activation function improved significantly after we increased $\alpha$ to the much larger value of 0.3. Leaky ReLU adjusts ReLU for the zero-gradient problem, adding a slant for negative values of ReLU. Although seismic data may seem complex, they contain a high degree of linearity (superimposed linear wavefields). Increasing the slope of Leaky ReLU pushes the activation function towards becoming linear. As this yields better results, it is likely that less complexity and more linearity in the model makes the model perform better.

**Choice of loss functions**

The two loss models tested were respectively mean square error (MSE) which is an L2-norm and the L1-norm minimum absolute error (MAE). It is evident from Fig.12 that MSE performed much worse than MAE. When employing MSE, the linear structure of the SI noise can be observed in the output of the network as a strong residual (Fig. 12a). Compared to MSE, MAE removes SI noise more efficiently (Fig. 12b).

**Choice of filter size**

The filter size increases the total number of parameters and thus has a direct impact on the execution time. Different filter sizes were tested as part of the design process for the network. A smaller filter size of 3×3 was tested in all convolutional layers (combination 1), but the performance of the trained network was found to be poorer compared to the use of larger filters. A reasonable choice seemed to be the use of 6×6 filters in the first two convolutional layers and 4×4 filters in the third convolutional layer (combination 2). Figure 13 shows an example of a SI-denoised shot gather for both filter combinations 1 and 2. It can be seen that by increasing the filter size in the first convolutional layers has increased the networks capability to discriminate between the SI-noise and the seismic signal to be recovered.

**FIELD DATA EXAMPLES**

To demonstrate the performance of the customized U-Net, we conducted a field data test. The data sets used here are the same as those used in previous section to test the performance of the AE.

An example of the denoising results obtained after proper training of the customized U-Net can be seen in Fig. 14, where the SI noise is coming from astern. The output, Fig. 14c, shows some residual SI noise, but it has weak amplitudes and with almost everything being attenuated. Box iii in Fig. 14c highlights a section where the high amplitude SI noise blends with the input data. Almost no residuals are left, as can be further illustrated in box vi in Fig. 14d containing the noise not removed. The underlying geology is well recreated as there is essentially no geological information left in the difference. Comparing sections with less noise, i.e. box ii in Fig. 14c and box v in Fig. 14d, show low-amplitude residual noise and virtually no geological loss. The customized U-Net does a good job at recreating the shallow events, i.e. box i in Fig. 14c and box iv in Fig. 14d, where almost no geological information is left in the difference. The frequency spectra of the ground truth and the denoised gather given in Fig. 14e, show apparently no deviation at any frequency. The loss plot in Fig. 14f shows a step-wise training loss which stabilizes at $10^{-4}$ after approximately 25 to 30 epochs. The validation loss is oscillating much until 15 epochs, but stabilizes at the same value as the training loss.

Figure 15 shows an example with SI noise coming from ahead (i.e. along the sail line direction in front of the recording vessel), where the network still performs rather well except in areas when the noise intercept with similar dips as the underlying geology.

The most challenging case is represented by SI noise coming from the side. However, as already mentioned, contracting companies will try to minimize such contributions during acquisition. For completeness, we have also investigated the performance of our network for this type of SI noise (see Appendix A).

**COMPARISON BETWEEN THE CUSTOMIZED U-NET AND AN INDUSTRY STANDARD ALGORITHM**

To handle the problem of SI contamination, the industry typically implements a processing flow consisting of multiple procedures which is an efficient but computer-intensive approach. In order to further test the quality of our customized U-Net in a strict way, we compare its performance on a noise contaminated stack from an area around 300km away from where we got our training dataset, with the result obtained employing an industry standard SI-denoising algorithm (Zhang and Wang 2015). This is essentially a sparse $\tau$–p transform, followed by random noise attenuation applied on common-p gathers. In such a comparison test, we can also test how robust and adaptive our network is when trained on one data set and employed on field data from a different geological area.

Figure 16 shows a noise contaminated stack from this new area. The denoising result of the proposed U-Net after proper stacking can be seen in Fig. 17. The overall denoising capability of the customized U-Net is rather good. It removes a significant amount of noise, while keeping almost all geology intact. There are patches with residual noise present in the denoised stack indicated by arrows in Fig. 17, but the remainder of the stack is well denoised. Figure 18 shows the denoising result employing an industry standard procedure. It is evident that the industry standard denoising performs better than the proposed network, especially in areas where the SI noise comes from the side. However, in regions where the SI noise comes from far ahead the industry algorithm performs poorer than the network (indicated by arrows in Fig. 18). Finally, if we compare the removed SI noise of the proposed network (Fig. 19) and the industry algorithm (Fig. 20), the latter shows no apparent geology and more noise removed in accordance with the previous observations.

An important aspect to mention in the comparison between industry denoising and the proposed network is the difference in data handling of the two. The customized U-Net is reading one shot at a time and denoising in real time. The industry standard is using a multi-shot approach to break the coherency of the data (Shen *et al.* 2019), thus giving the industry denoising an advantage over the neural network in denoising accuracy. This could in principle be done for the network as well, but will not be optimal since we want to keep the advantages of the network when it comes to fast-track processing of large data volumes as well as in real-time processing and quality control onboard seismic vessels during acquisition.

The efficiency of denoising algorithms are important to the industry. The less computing time needed for denoising, the more money saved. After years of research, SI noise can be almost perfectly attenuated by standard commercial algorithm. However, it is computationally heavy and time-consuming since a processing flow consisting of multiple procedures need to be implemented. The execution times of the proposed network and standard SI-denoising industry algorithm are compared in Table 2.

We can see that the customized U-Net is very efficient when fully trained. In general, the results shown in this paper cannot compete with such industry standard for now, but the proposed network performs orders of magnitude better than industry standard denoising with respect to time.

**DISCUSSION**

Neural networks have become increasingly popular during the last 5 to 10 years within many areas of technology. Although employed to solve various problems within conventional imaging and classification, less studies have been carried out in the field of seismic data processing. A neural network-based seismic denoising approach has the potential of real time processing. Thus, it opens up for the possibility to test whether data have to be re-shot during acquisition. The data flow in continuously and real time denoising by using neural network makes it easier to efficiently create rough estimates to whether the data are too contaminated or not. It is less common for contractor companies to actually re-shoot nowadays, but it may still prove a valuable tool for quality control geophysicists on board. It may also be applied on data that are only lightly affected by SI noise, thus saving significant computer time. A typical marine survey may contain around 10 million source-cable shot gathers. If all of these were to be run on a single GPU at 4s each, this would take more than one year. However, it would take only 2 days by the trained network.

The customized U-Net opens up for more efficient and robust quality control of seismic processing possibly in real time. It may then be possible to check whether any interesting features exist in the data and if it is worth spending time and resources on processing. If data with no important information can be omitted easily, the contractor companies might increase their efficiency and spend more time on important data. A hybrid approach where a CNN is trained on a subset of 3D data being industry processed and then employed on the remaining part of the data set may lead to almost real-time processing. Another example would be to train on vintage data and then denoise in real time when new data come in. As such, neural network driven seismic data processing could become a very valuable approach in the not too distant future.

**CONCLUSIONS**

When multiple marine seismic surveys are carried out simultaneously, SI noise becomes a problem. In this paper we have studied a deep learning-based approach to attenuate SI noise using CNNs. An initial attempt to apply the commonly used AE (autoencoder) to solve the seismic denoising problem proved unfeasible. Thus, we have designed a customized U-Net according to the special characters of both seismic signal and noise.

The proposed network was trained on numerically blended SI noise contaminated shot gathers, and then its performance was verified on a set of test data. The SI noise used in this study is unusually strong which therefore creates a more complex problem than what normally encountered during seismic acquisition. Thus, the performance of the proposed network is determined for a highly realistic acquisition case.

The customized U-Net performed overall well with a high percentage of the SI noise being attenuated, except when the noise came from the side. This latter case is known to be challenging and the contractor companies try to minimize its influence during acquisition. In a small supplementary study, we demonstrated that also such SI noise can be quite efficiently removed if we introduced a deeper version of our proposed customized network.

To further investigate the performance of our network, we compared our results with the results obtained by using a standard commercial algorithm. Even though the results from our network still cannot compete completely with a standard commercial algorithm, it has been demonstrated that attenuating SI noise with neural networks is feasible. The customised U-Net has great advantages in computational efficiency. It required only 7.5 hours to train on a modern GPU with powerful computational ability. Once finished training it takes only 0.02s to denoise each shot gather, which is significantly faster than any existing industry denoising technique. Another important issue is that the network can be applied shot by shot. This is a further advantage over commercial algorithms, which typically require a multi-shot input to break the coherency of the noise. This makes a network approach suitable for fast-track processing of large data volumes as well as for real-time processing and quality control onboard seismic vessels during acquisition.

## ACKNOWLEDGEMENTS

The authors thank CGG Data Library for permission to use the data shown in this article. The CNNs studied in this paper were implemented using Keras developed by Google.

## DATA AVAILABILITY STATEMENT

Seismic data used in this paper are the property of CGG.

**APPENDIX A: THE CHALLENGE OF SEISMIC INTERFERENCE NOISE FROM THE SIDE**

As discussed in the main body of the paper, contracting companies try to minimize the influence of SI noise from the side. However, in practice, some contributions of such SI noise are inevitable. For completeness, we will therefore also discuss this special case.

In Fig. A-1b the case of high-amplitude SI noise coming from the side is illustrated. This type of SI noise appears with much curvature and is kinematically similar to reflection data. Thus, it represents a challenge for seismic data processing. Application of our customized network gave the result shown in Fig. A-1c. It can be seen that the network failed in areas where the underlying geology is completely masked by the SI-noise creating a pattern of blank zones (see arrows). Note that the result shown could possibly have improved if we had trained the network with more examples of this specific type of SI noise.

Realizing that field data sets may often be limited in diversity, we investigated next the idea of increasing the depth of our customized U-Net. The idea is that such an extended network may be able to extract more features from the data available. Thus, we modified the current version of our network with two more layers. Application to the same data as shown in Fig. A-1, gave the result shown in Fig. A-2. It can be easily seen that the denoising result has improved considerably and with almost no residual SI-noise (Fig. A-2a). This is also confirmed from inspection of the difference plot in Fig. A-2b. In this case, the training time of the network is about 9 hours on the same GPU while the testing time is very close to before.

# REFERENCES

Abadi M., Barham P., Chen J., Chen Z., Davis A., Dean J., Devin M., Ghemawat S., Irving G., Isard M., Kudlur M., Levenberg J., Monga R., Moore S., Murray D. G., Steiner B., Tucker P., Vasudevan V., Warden P., Wicke M., Yu Y. and Zheng X. 2016. Tensorflow: A system for large-scale machine learning. Proceedings of the 12th USENIX Symposium on Operating Systems Design and Implementation, Savannah, GA, USA, 265-283.

Akbulut K., Saeland O., Farmer P. and Curtis T. 1984. Suppression of seismic interference noise on Gulf of Mexico data. 54th SEG Annual Meeting, Atlanta, USA, Expanded Abstracts, 527–529.

Baardman R. 2018. Classification and Suppression of Blending Noise Using CNN. First EAGE/PESGB Workshop Machine Learning, London, UK, Extended Abstracts, ML 11.

Baardman R. and Tsingas C. 2019. Classification and Suppression of Blending Noise Using Convolutional Neural Networks. SPE Middle East Oil and Gas Show and Conference, Manama, Bahrain, Conference Papers, SPE-194731-MS.

Deng L. and Yu D. 2014. Deep Learning: Methods and Applications. Foundations and Trends in Signal Processing 7, (3), 197-387.

Dumoulin V. and Visin F. 2018. A guide to convolution arithmetic for deep learning. arXiv, 1603.07285.

Elboth T., Presterud I. V. and Hermansen D. 2010. Time-frequency seismic data de-noising. Geophysical Prospecting 58, (3), 441-453.

Fukushima K. 2007. Neocognitron. Scholarpedia 2, (1), 1717.

Gondara L. 2016. Medical image denoising using convolutional denoising autoencoders. 16th

IEEE International Conference on Data Mining Workshops (ICDMW), Barcelona, Spain, Extended Abstracts, 241-246.

Goodfellow I., Bengio Y. and Courville A. 2016. Deep learning. MIT press.

Gulunay N. and Pattberg D. 2001(a). Seismic crew interference and prestack random noise attenuation on 3-D marine seismic data. 63rd EAGE Conference and Exhibition, Amsterdam, The Netherlands, Extended Abstracts, A-13.

Gulunay N. and Pattberg D. 2001(b). Seismic interference noise removal. 71st SEG Meeting, San Antonio, Texas, US, Expanded Abstracts, 1989-1992.

Gulunay N., Magesan M. and Baldock S. 2004. Seismic interference noise attenuation. 74th SEG Technical Program, Denver, Colorado, US, Expanded Abstracts, 1973-1976.

Gulunay N. 2008. Two different algorithms for seismic interference noise attenuation. The Leading Edge 27, (2), 176-181.

Guo J. and Lin D. 2003. High-amplitude noise attenuation. 73rd SEG Meeting, Dallas, Texas, US, Expanded Abstracts, 1893-1896.

Gwardys G. 2016. Convolutional Neural Networks backpropagation: from intuition to derivation, https://grzegorzgwardys.wordpress.com/, accessed 19 June 2016.

He K., Zhang X., Ren S. and Sun J. 2016. Deep residual learning for image recognition. Proceedings of the 29th IEEE Conference on Computer Vision and Pattern Recognition, Las Vegas, NV, USA, 770-778.

Hubel D. H. and Wiesel T. N. 1968. Receptive fields and functional architecture of monkey striate cortex. The Journal of Physiology 195, (1), 215-243.

Jansen S., Elboth T. and Sanchis C. 2013. Two Seismic Interference Attenuation Methods

Based on Automatic Detection of Seismic Interference Moveout. 75th EAGE Annual Meeting, London, UK, Extended Abstracts, We 14 15.

Kommedal J. H., Semb P. H. and Manning T. 2007. A case of SI attenuation in 4D seismic data recorded with a permanently installed array. Geophysics 72, (3), Q11-Q14.

Laurain R., Ruiz-Lopez F. and Eidsvig S. 2014. Improving acquisition efficiency by managing and modelling seismic interference. First Break 32, (11), 79-85.

Li F.-F., Johnson J. and Young S. 2017. Lecture 11: Detection and segmentation. CS231n: Convolutional Neural Networks for Visual Recognition, Stanford.

Liu D., Wang W., Chen W., Wang X., Zhou Y. and Shi Z. 2018. Random noise suppression in seismic data: What can deep learning do?. 88th SEG Meeting, Anaheim, CA, US, Expanded Abstracts, 2016-2020.

Ma J. 2018. Deep Learning for Attenuating Random and Coherence Noise Simultaneously. 80th EAGE International Conference and Exhibition, Copenhagen, Denmark, Extended Abstracts, Tu P7 07.

Maas A. L., Hannun A. Y. and Ng A. Y. 2013. Rectifier nonlinearities improve neural network acoustic models. Proceedings of the 30th International Conference on Machine Learning, Atlanta, US, 3.

Mandelli S., Borra F., Lipari V., Bestagini P., Sarti A. and Tubaro S. 2018. Seismic data interpolation through convolutional autoencoder. 88th SEG Meeting, Anaheim, CA, US, Expanded Abstracts, 4101-4105.

Mandelli S., Lipari V., Bestagini P. and Tubaro S. 2019. Interpolation and Denoising of Seismic Data using Convolutional Neural Networks. arXiv, 1901.07927.

Matsugu M., Mori K., Mitari Y. and Kaneda Y. 2003. Subject independent facial expression recognition with robust face detection using a convolutional neural network. Neural Networks 16, (5-6), 555-559.

Mehta P., Bukov M., Wang C. H., Day A. G. R., Richardson C., Fisher C. K. and Schwab D. J. 2019. A high-bias, low-variance introduction to machine learning for physicists. Physics Reports, 810, 1-124.

Mitchell T. M. 1997. Machine Learning. McGraw-Hill Higher Education.

Paszke A., Gross S., Chintala S., Chanan G., Yang E., DeVito Z., Lin Z., Desmaison A., Antiga L. and Lerer A. 2017. Automatic differentiation in PyTorch. 31st Conference on Neural Information Processing Systems, Long Beach, California, USA.

Qian F., Yin M., Liu X. Y., Wang Y. J., Liu C. and Hu G. M. 2018. Unsupervised seismic facies analysis via deep convolutional autoencoders. Geophysics 83, (3), A39-A43.

Rentsch S., Holicki M. E., Kamil Y. I., Robertsson J. O. A. and Vassallo M. 2014. How to teach a neural network to identify seismic interference. 76th EAGE Conference and Exhibition, Amsterdam, The Netherlands, Extended Abstracts, Th ELI1 07.

Ronneberger O., Fischer P. and Brox T. 2015. U-net: Convolutional networks for biomedical image segmentation. Springer International Conference on Medical image computing and computer-assisted intervention, Cham, Conference Papers, 234-241.

Shen H., Elboth T., Tao C., Tian G., Wang H., Qiu L. and Zhou J. 2019. Using data regrouping methods to attenuate shot-to-shot coherent interference noise in marine seismic data. Earth and Space Science 6.

Slang S., Sun J., Elboth T., McDonald S. and Gelius L. J. 2019. Using Convolutional Neural

Networks for Denoising and Deblending of Marine Seismic Data. 81st EAGE Conference and Exhibition, London, UK, Extended Abstracts, Tu R06 05.

Tieleman T. and Hinton G. 2012. Divide the gradient by a running average of its recent magnitude: COURSERA. Neural networks for machine learning 4.2, 26-31.

Waldeland A. U., Jensen A. C., Gelius L. J. and Solberg A. H. S. 2018. Convolutional neural networks for automated seismic interpretation. The Leading Edge 37, 529-537.

Wang B. F., Zhang N., Lu W. K., Zhang P. and Geng J. H. 2018a. Seismic Data Interpolation Using Deep Learning Based Residual Networks. 80th EAGE Conference and Exhibition, Copenhagen, Denmark, Extended Abstracts, Th A15 14.

Wang B. F., Zhang N., Lu W. K. and Wang J. 2018b. Deep learning-based seismic data interpolation: A preliminary result. Geophysics 84, (1), V11-V20.

Wang H., Liu G., Hinz C. E. and Snyder F. F. C. 1989. Attenuation of marine coherent noise. 59th SEG Annual Meeting, Dallas, Texas, US, Expanded Abstracts, 1112-1114.

Willmott C. J. and Matsuura K. 2005. Advantages of the mean absolute error (MAE) over the root mean square error (RMSE) in assessing average model performance. Climate research 30, (1), 79-82.

Wu H., Zhang B., Li F. and Liu N. 2019. Semi-automatic first arrival picking of micro-seismic events by using pixel-wise convolutional image segmentation method. Geophysics 84, (3), V143-V155.

Xie J., Xu L. and Chen E. 2012. Image denoising and inpainting with deep neural networks. Advances in neural information processing systems 25, 341-349.

Xiong W., Ji X., Ma Y., Wang Y., AlBinHassan N. M., Ali M. N. and Luo Y. 2018. Seismic fault

detection with convolutional neural network. Geophysics 83, (5), O97-O103.

Zhang Z. and Wang P. 2015. Seismic interference noise attenuation based on sparse inversion. 85th SEG Annual Meeting, New Orleans, Louisiana, US, Expanded Abstracts, 4662-4666.

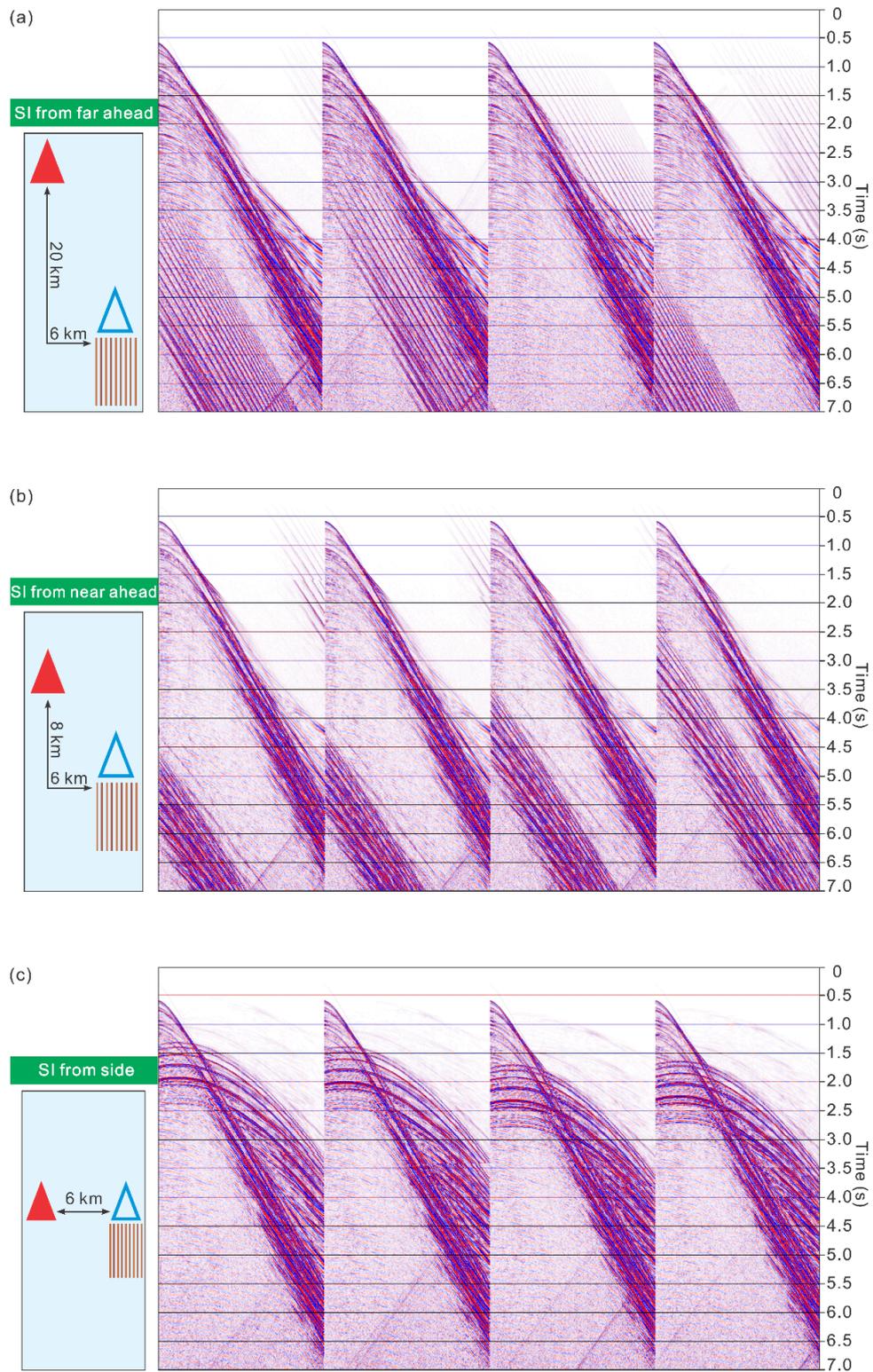

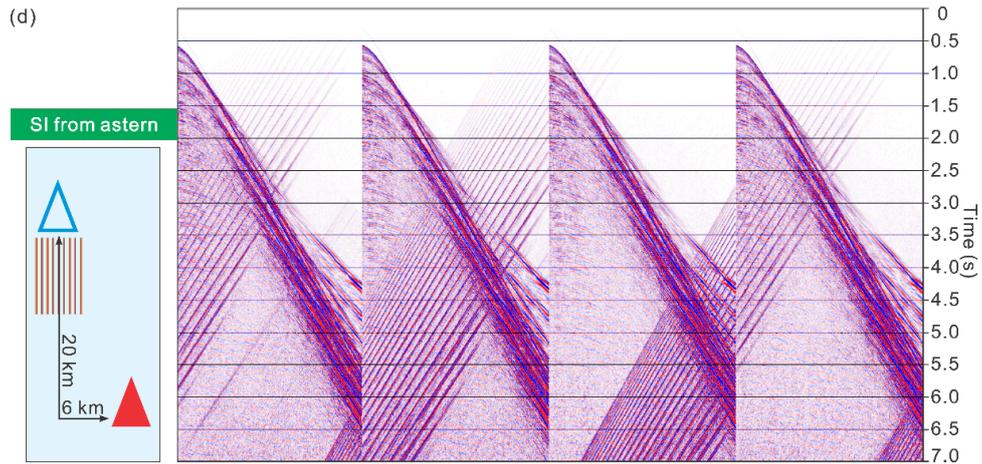

**Figure 1** Example of SI noise from: (a) far ahead, (b) near ahead, (c) the side, and (d) astern.

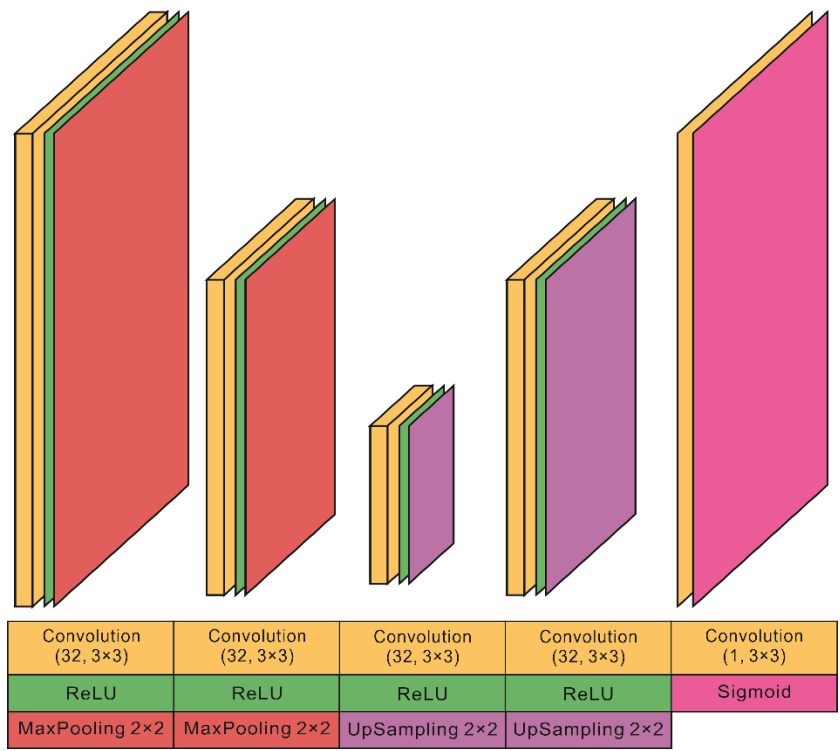

**Figure 2** A schematic of the AE for SI-denoising. Yellow rectangles represent convolutional operation. The initial value in brackets (i.e. 32 or 1) indicates the number of filters, followed by 3×3 which is the filter size. Green rectangles represent the activation function 'ReLU'. Red rectangles represent the max pooling operation where 2×2 is the pool size (i.e. data are being downscaled to half size in both spatial dimensions). Purple rectangles represent the upsampling operation with 2×2 representing the upsampling factors.

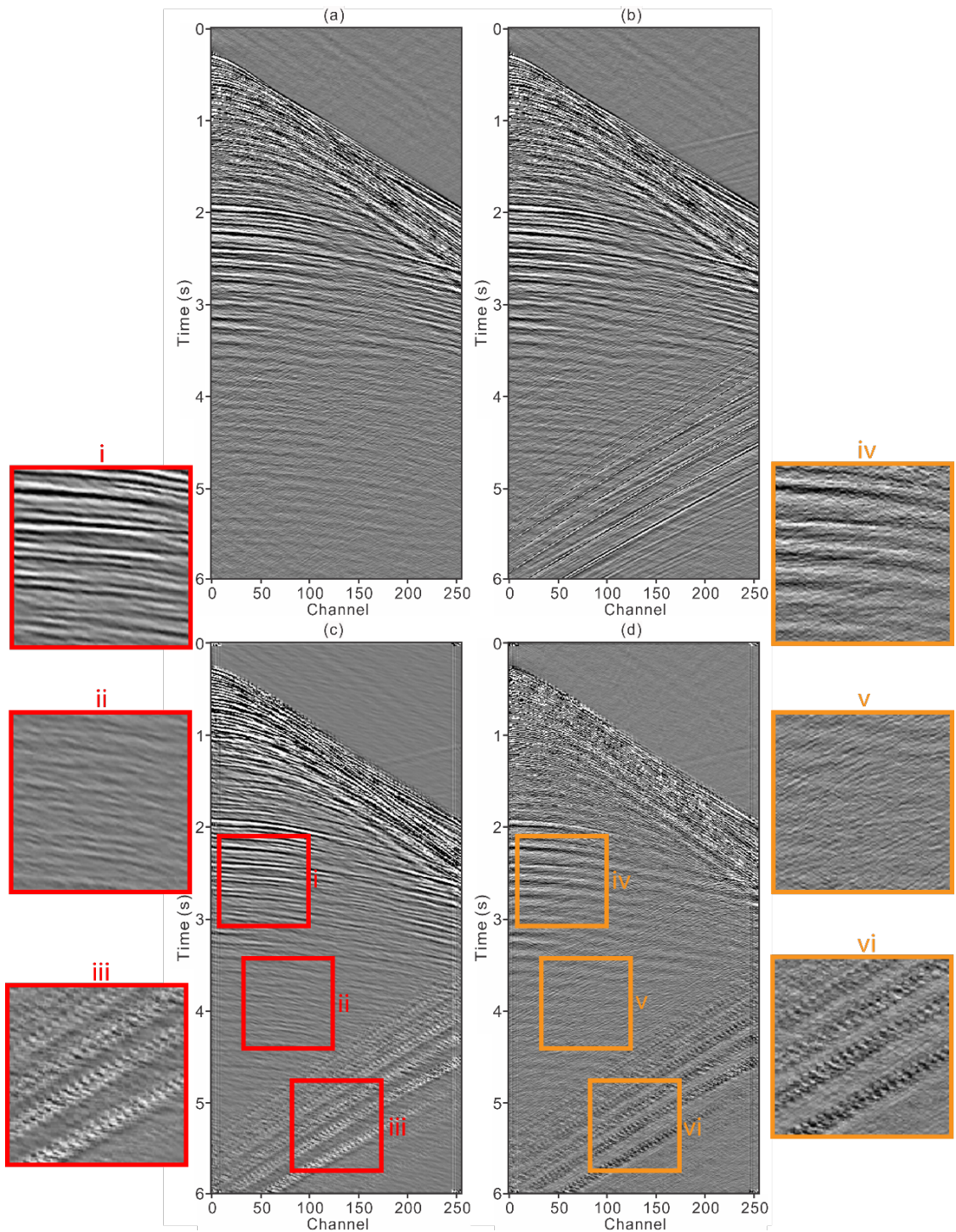

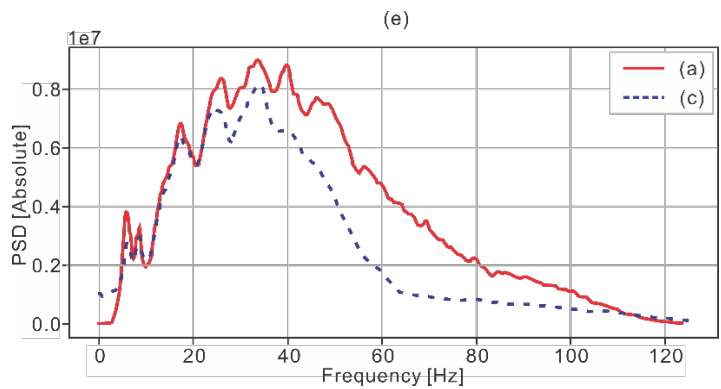

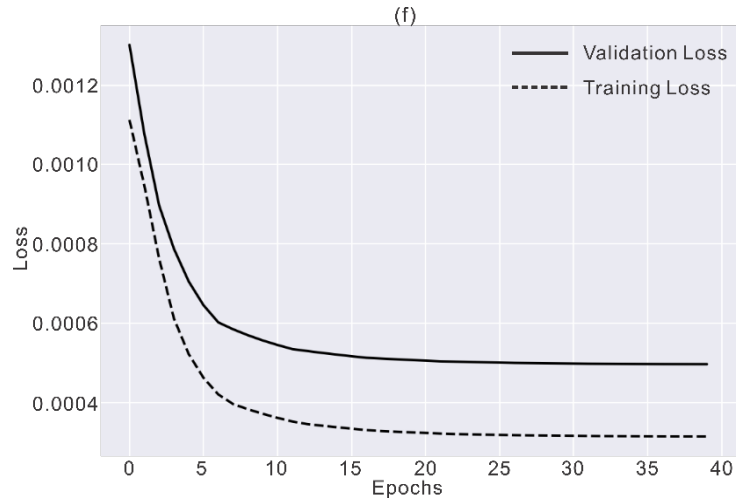

**Figure 3** An example of SI-denoising using the AE in Fig.2 with SI noise coming from astern: (a) ground truth/clean shot gather, (b) SI noise contaminated shot gather, (c) denoised shot gather, (d) difference between ground truth and denoised shot gather, (e) frequency spectra of ground truth and denoised shot gather, and (f) training and validation loss. Boxes i, ii, ..., vi show zoomed sections of interest. The red sections mark locations in the denoised data (c), while the orange sections mark corresponding areas in the difference (d).

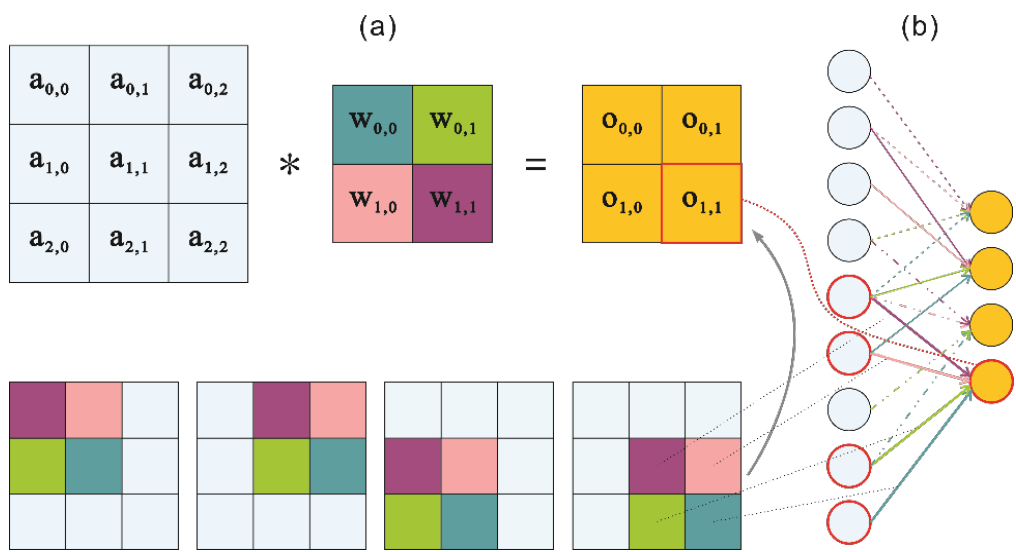

**Figure 4** A schematic representation of feedforward in CNNs (adapted from Gwardys 2016). (a) 2D convolution operation as an alternative representation of a (b) connected network. Filter weights are highlighted in four different colors.

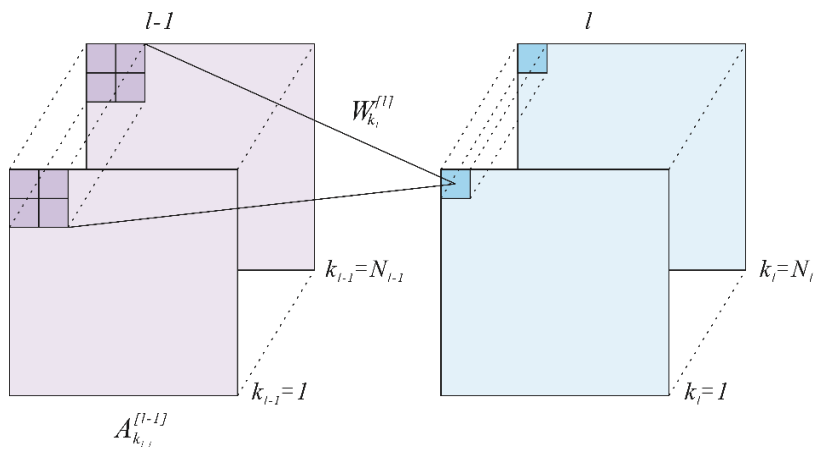

**Figure 5** Visualization outlining the convolution process of going from one arbitrary layer $l-1$ to layer $l$. The matrix $W_{k_l}^{[l]}$ represents the $k_l$'th kernel spanning all activations in layer $l-1$ producing each of the $k_l = 1, 2, ..., N_l$ feature maps in layer $l$.

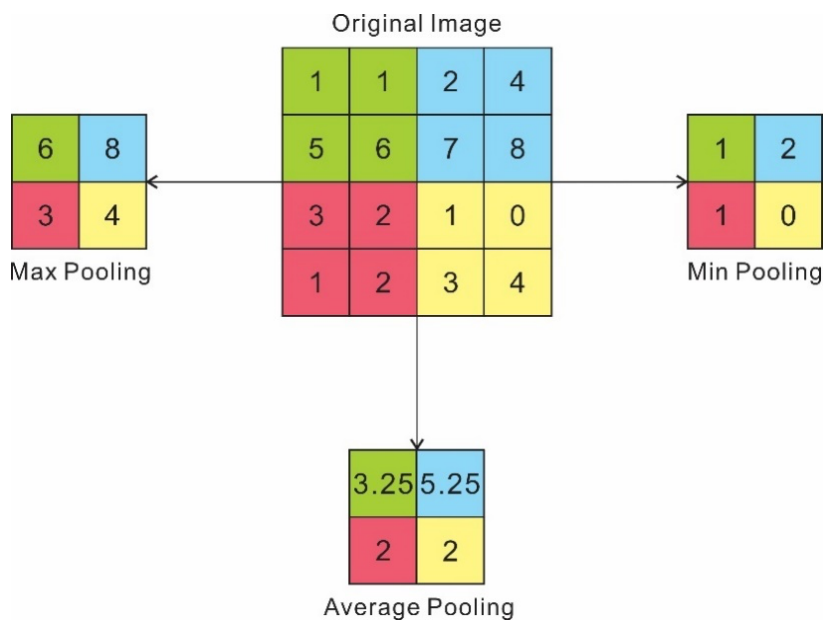

**Figure 6** Visual examples of 2×2 pooling of different kinds.

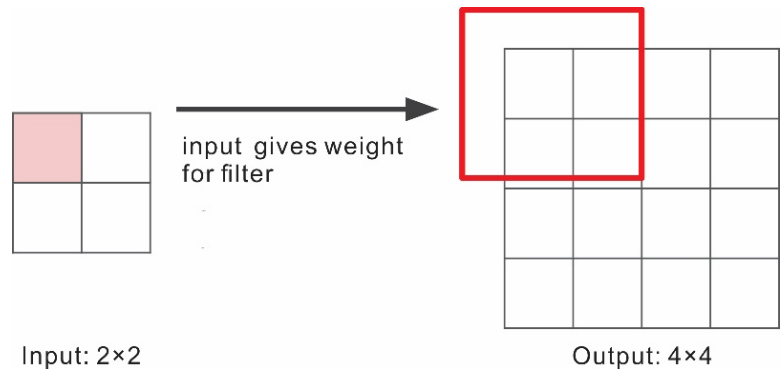

**Figure 7** Figure illustrating the process of upsampling with a factor of 2 (Li *et al.* 2017).

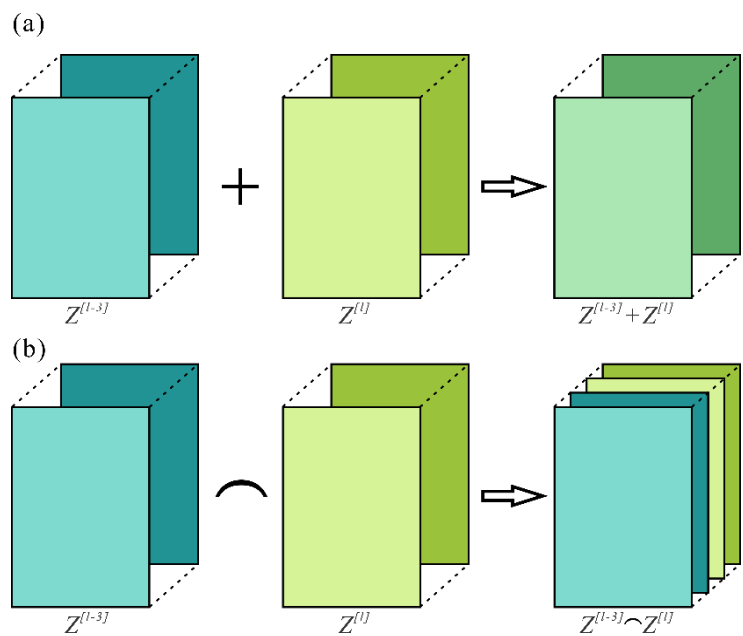

**Figure 8** Schematics of the operations of (a) element-wise summation and (b) concatenation. $Z^{[l-3]}$ represents a collection of the feature maps from layer $l-3$. $Z^{[l]}$ represents a collection of the feature maps from layer $l$.

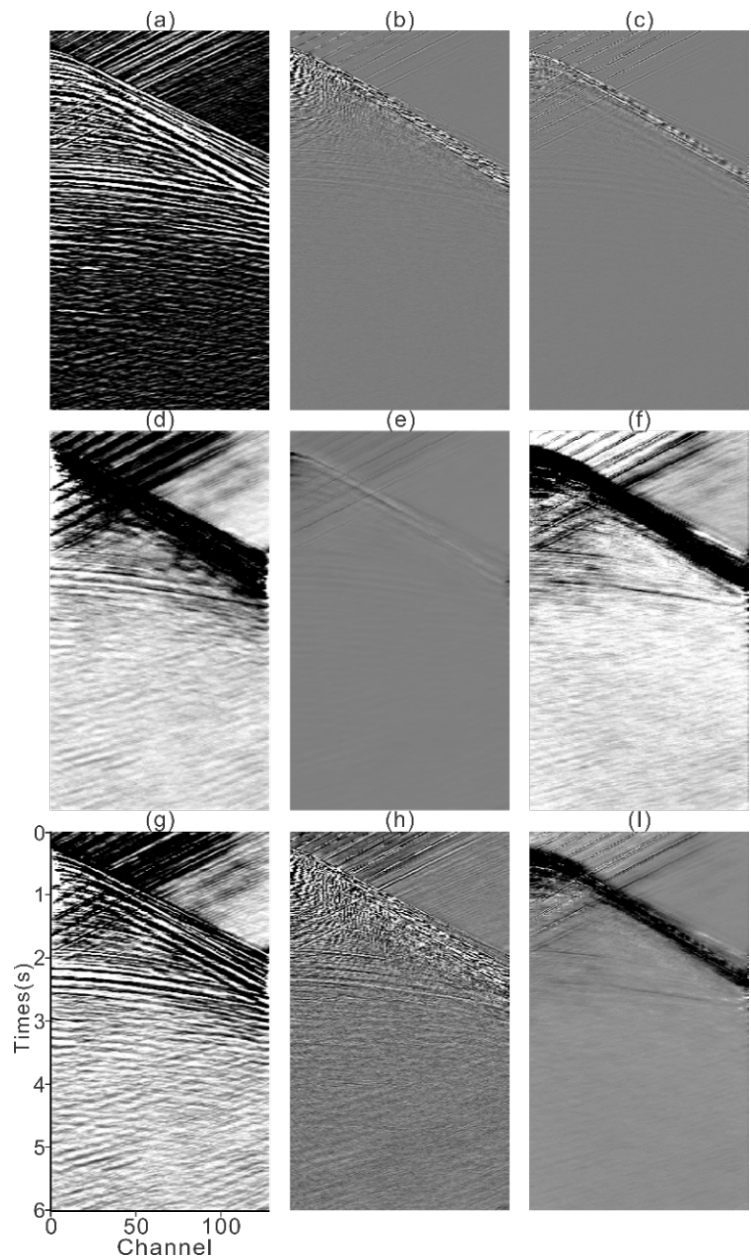

**Figure 9** An example of applying element-wise summation of the feature maps from a general layer and a deeper layer of the proposed network. (a), (b), (c) are selected examples of the feature maps of the first convolutional layer. (d), (e), (f) are the corresponding feature maps from the second upsampling layer (a deeper layer). (g), (h), (i) are the element-wise summation results.

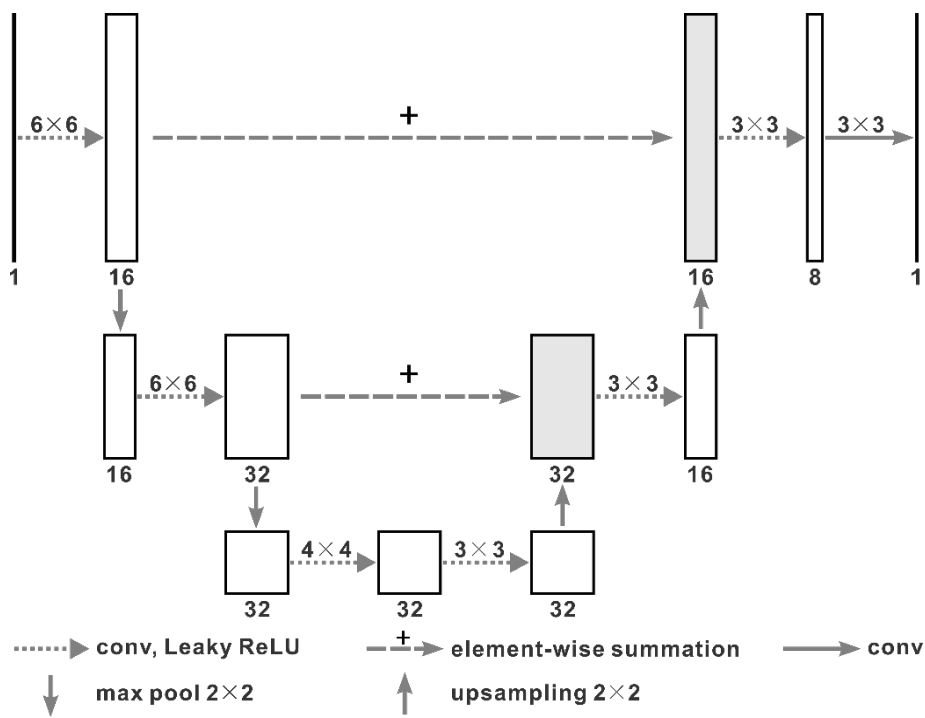

**Figure 10** A schematic of the customized U-Net for SI-denoising. The numbers 6×6, 4×4, and 3×3 indicate the filter size of the convolutional operation. Each rectangle represents a collection of feature maps from the previous operation, and the numbers below (1, 8, 16 and 32) represent the number of feature maps.

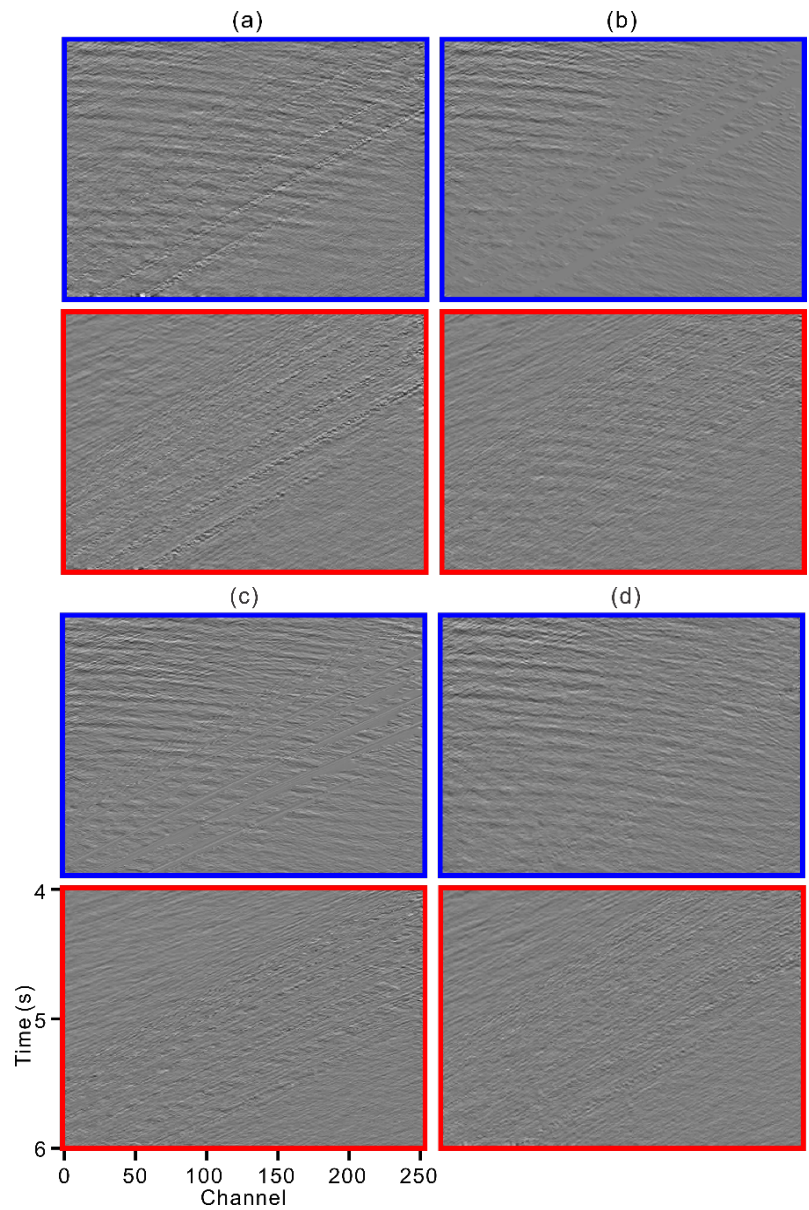

**Figure 11** Performance of the three different activation functions tested in the proposed network: (a) TanH, (b) ReLU, (c) Leaky ReLU with $\alpha = 0.01$, and (d) Leaky ReLU with $\alpha = 0.3$. The blue and red boxes represent respectively denoised result and difference between ground truth and denoised result.

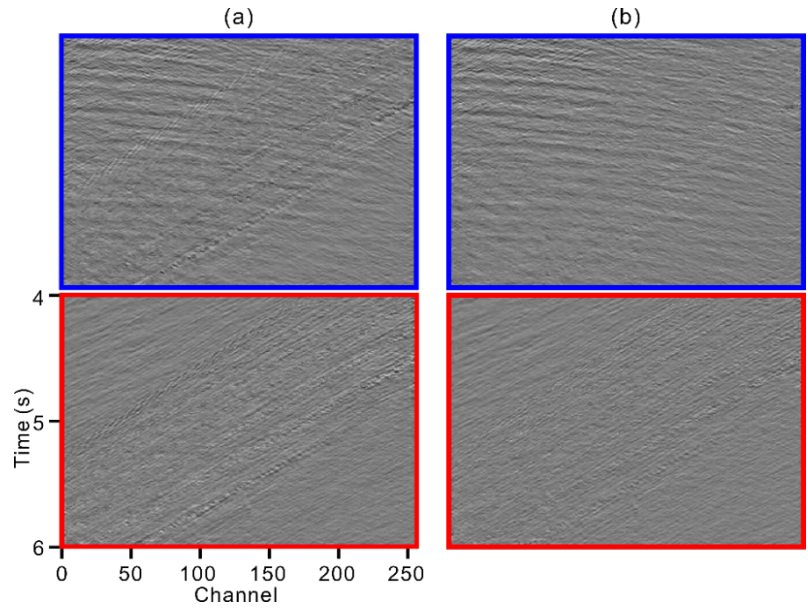

**Figure 12** Performance of the two different loss functions tested: (a) MSE and (b) MAE. The blue and red boxes represent respectively denoised result and difference between ground truth and denoised result.

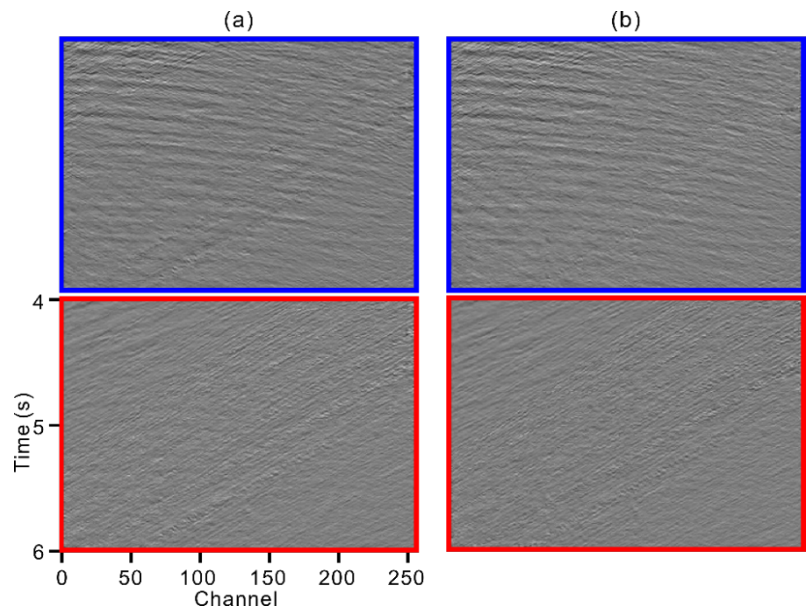

**Figure 13** Effect of varying filter size on the SI-denoised result: (a) filter combination 1 and (b) filter combination 2. The blue and red boxes represent respectively denoised result and difference between ground truth and denoised result

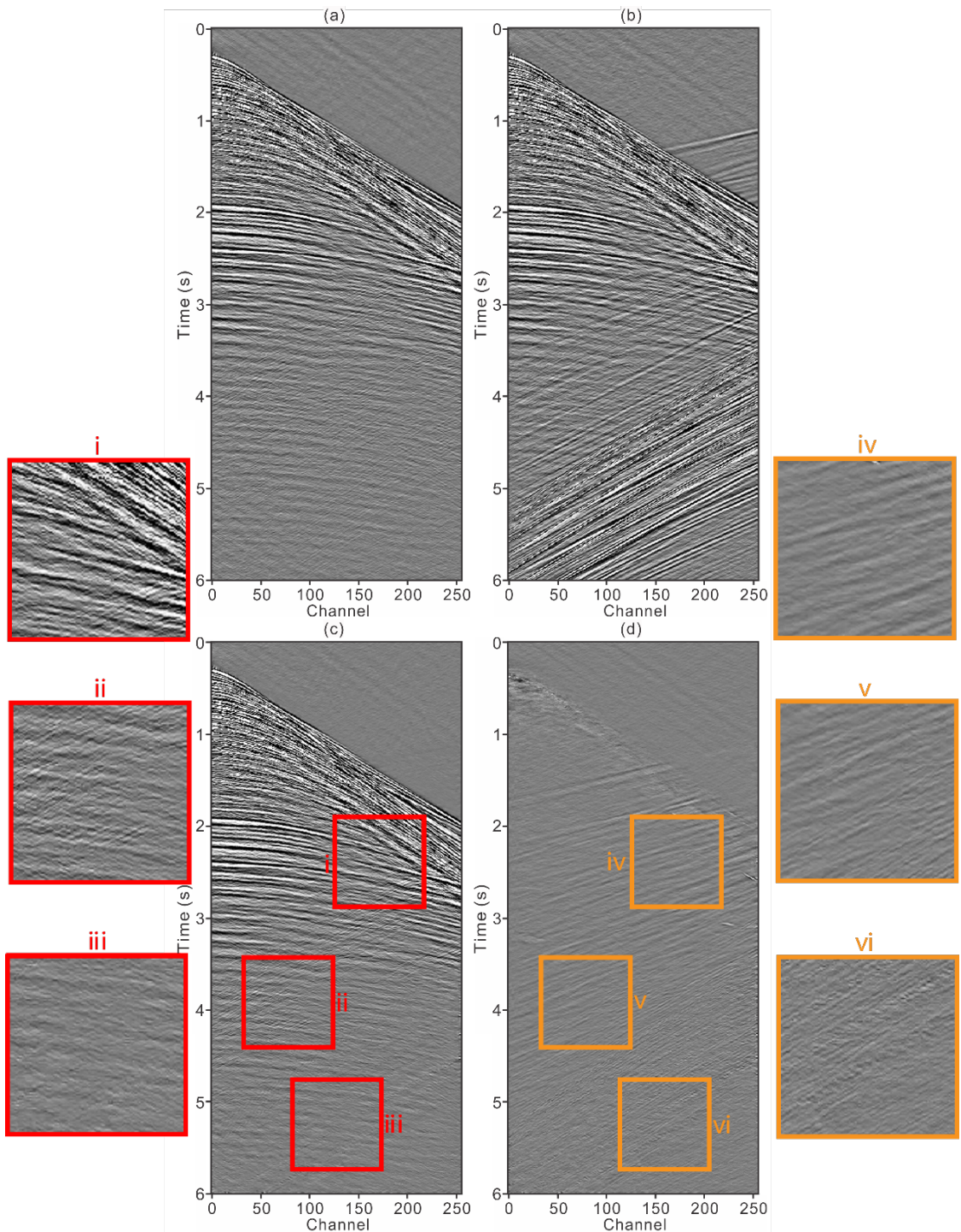

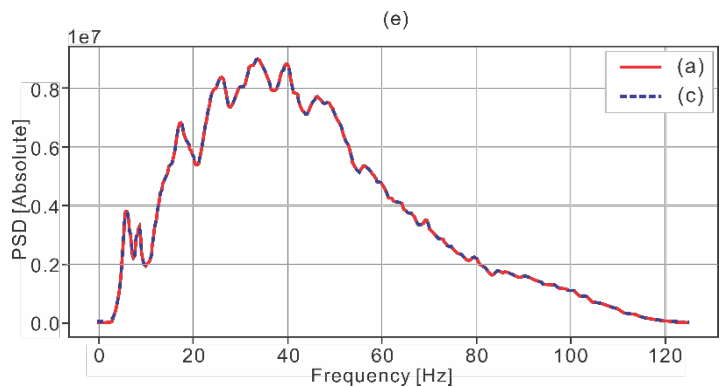

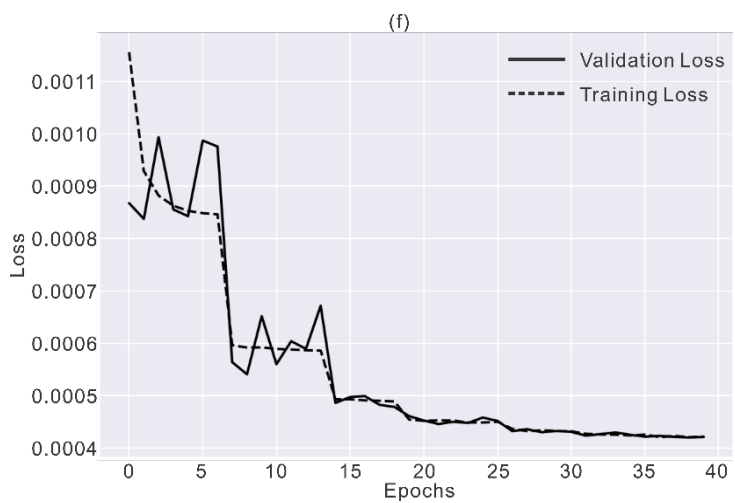

**Figure 14** An example of SI-denoising by the customized U-Net with SI noise coming from astern: (a) ground truth/clean shot gather, (b) SI-noise contaminated shot gather, (c) denoised shot gather, (d) difference between ground truth and denoised shot gather, (e) frequency spectra of ground truth and denoised shot gather, and (f) training and validation loss. Boxes i, ii, ..., vi show zoomed sections of interest. The red sections mark locations in the denoised data (c), while the orange sections mark corresponding areas in the difference (d).

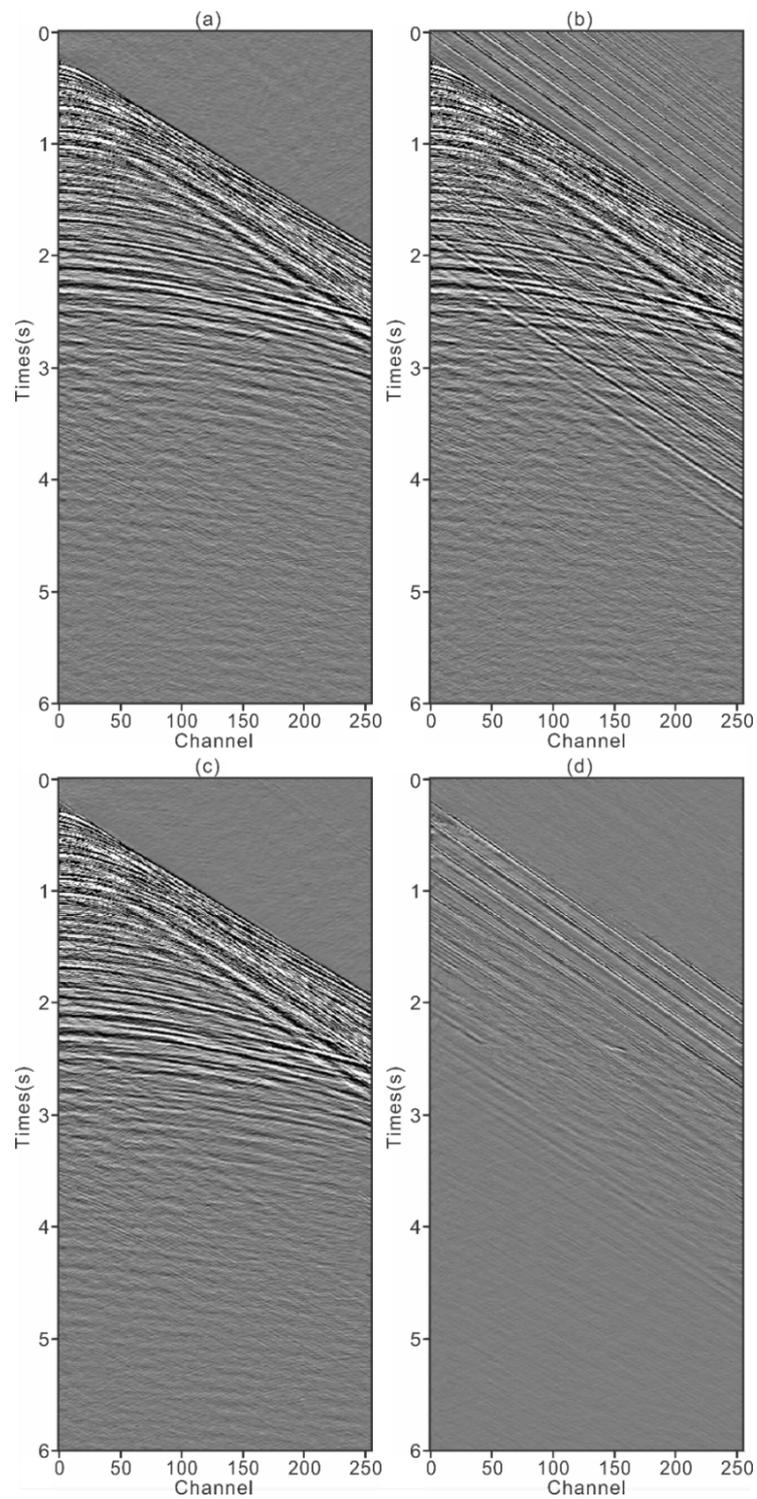

**Figure 15** An example of SI-denoising by the customized U-Net with SI noise coming from ahead: (a) ground truth/clean shot gather, (b) SI-noise contaminated shot gather, (c) denoised shot gather, and (d) difference between ground truth and denoised shot gather.

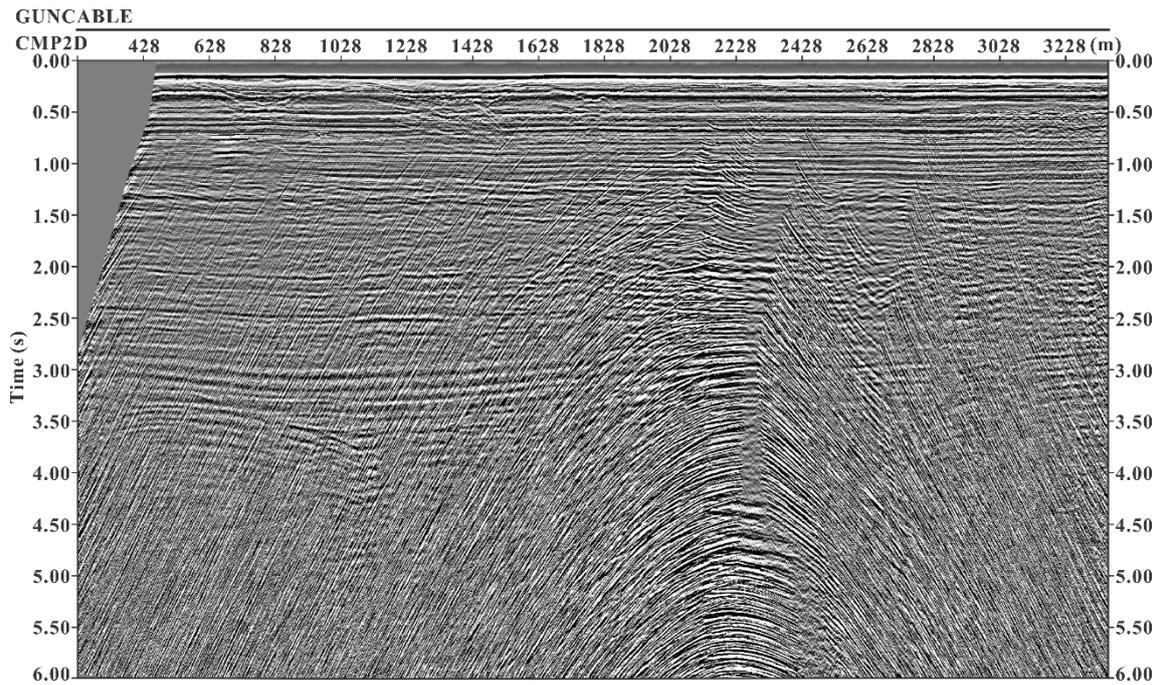

**Figure 16** SI-noise contaminated stack.

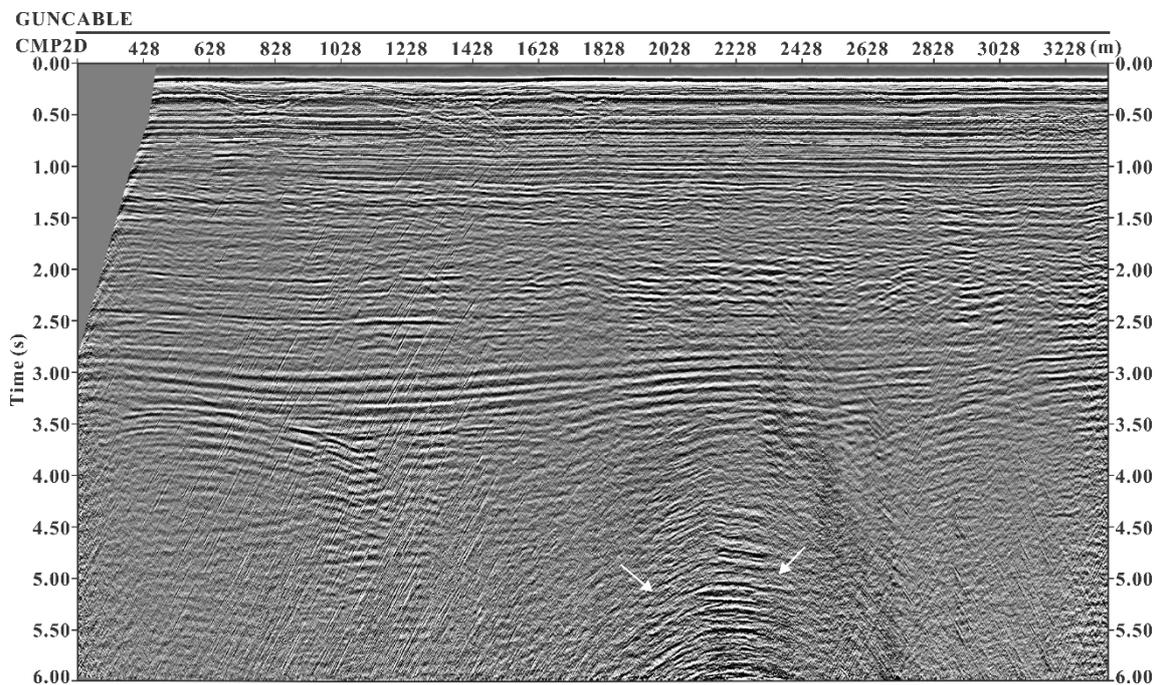

**Figure 17** Denoised stack after application of the trained network.

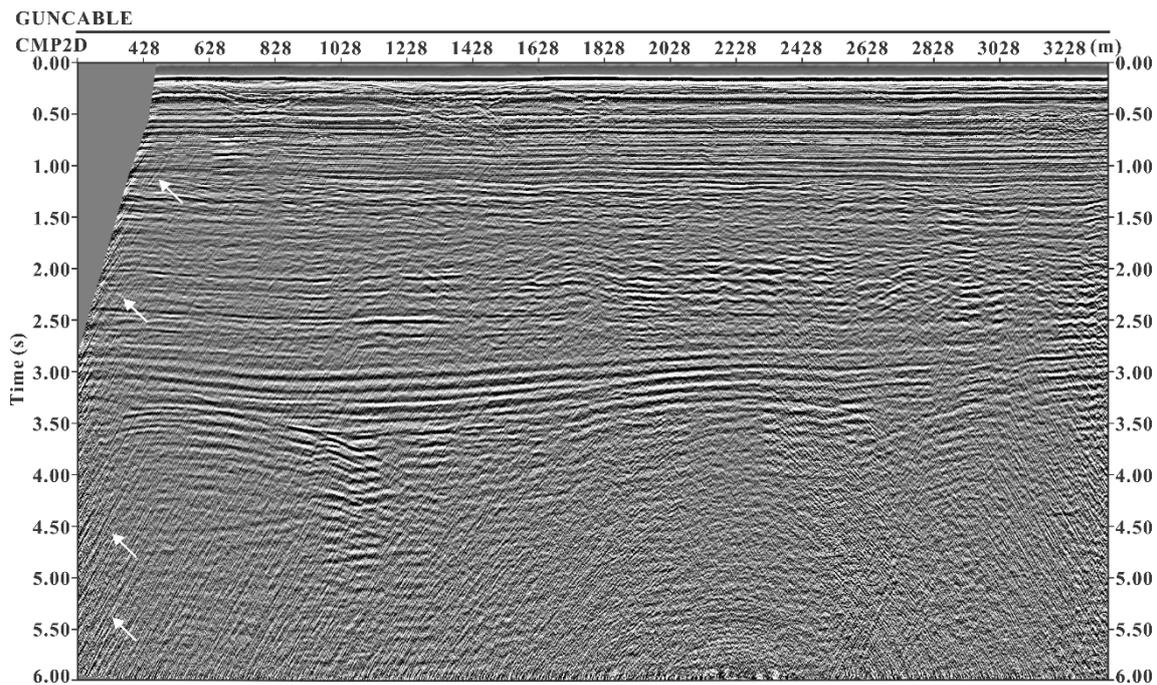

**Figure 18** Denoised stack employing an industry standard SI-denoising method.

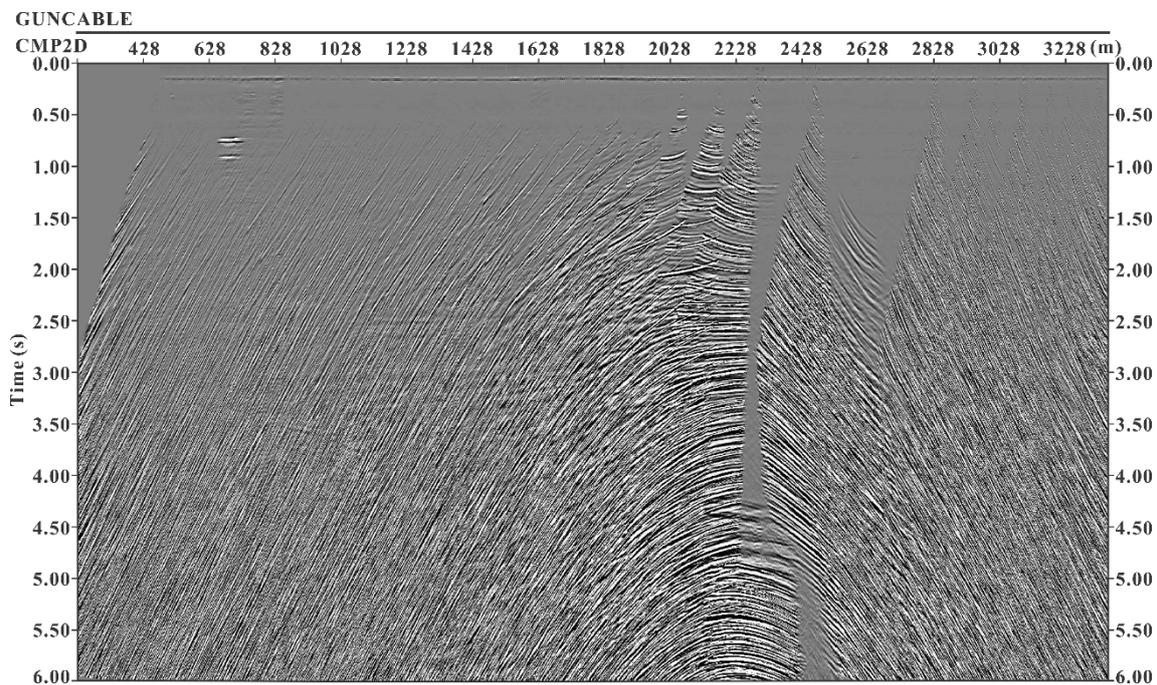

**Figure 19** Removed SI noise by the proposed network.

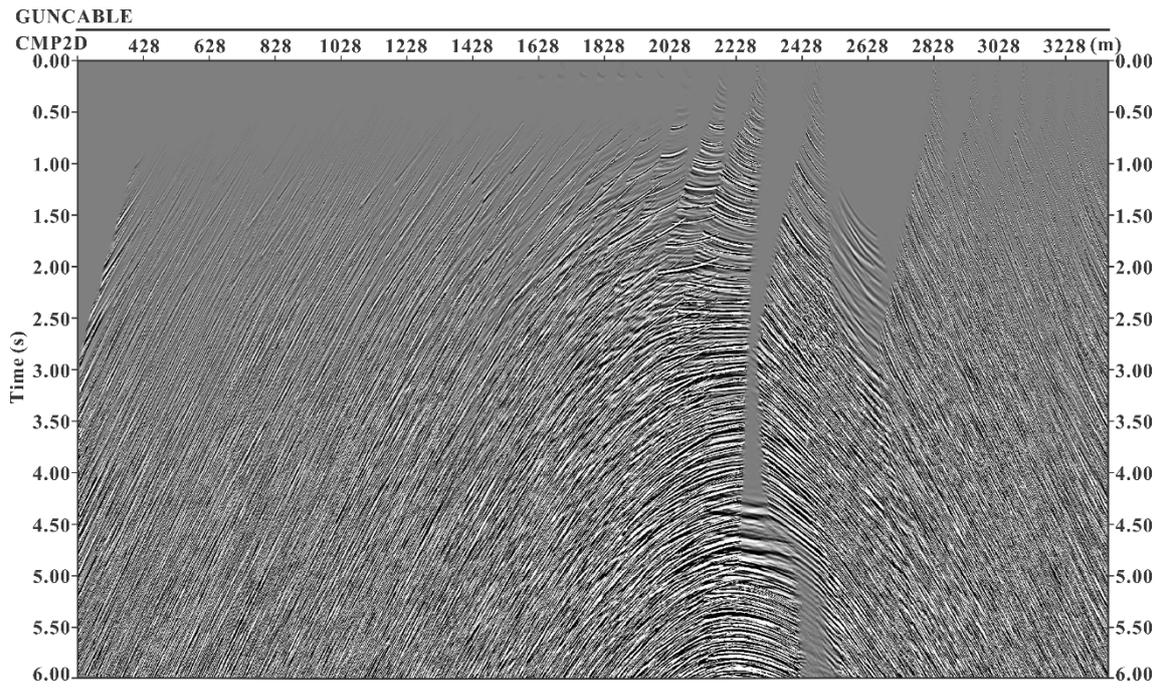

**Figure 20** Removed SI noise by industry standard method.

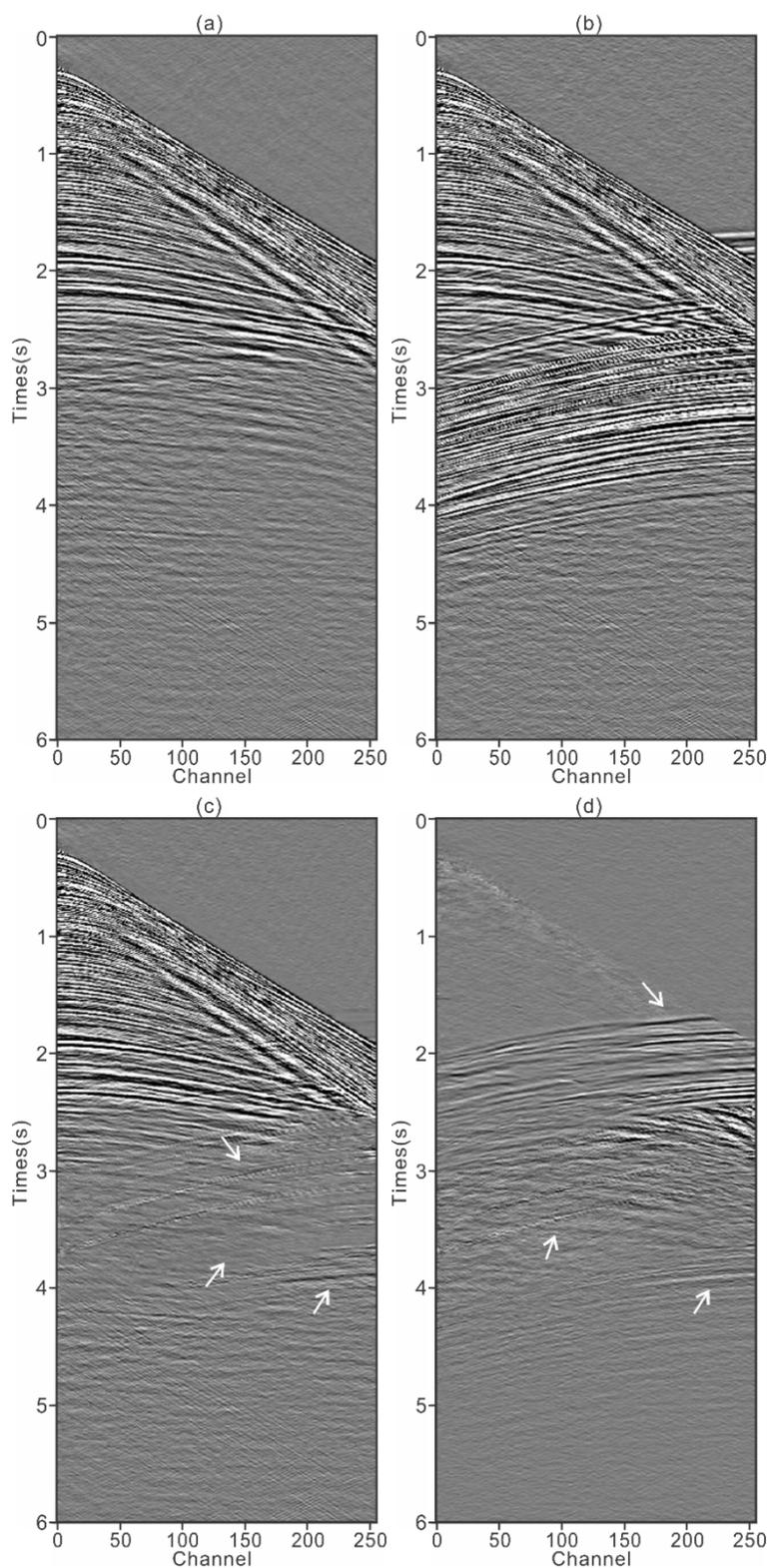

**Figure A-1** An example of SI-denoising by the customized U-Net with SI noise coming from the side: (a) ground truth/clean shot gather, (b) SI noise contaminated shot gather, (c) denoised shot gather, and (d) difference between ground truth and denoised shot gather.

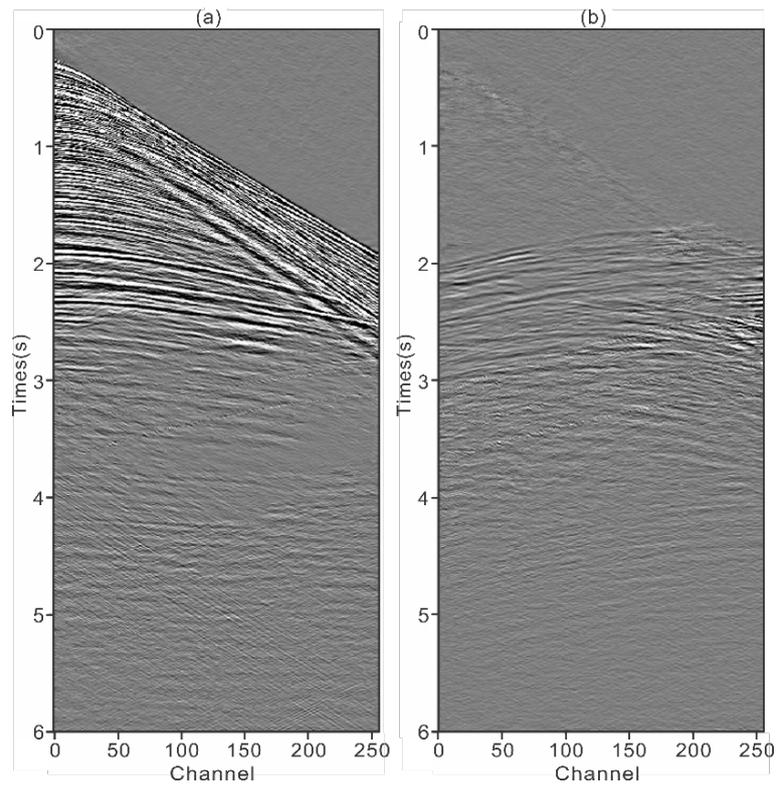

**Figure A-2** Results obtained employing a two-layer deeper network: (a) denoised shot gather and (b) difference between ground truth and denoised shot gather. Ground truth and SI noise contaminated shot gather used here are the same as in Fig. A-1.

Table 1 Differences between seismic data and conventional images.

| Differences | Seismic data | Conventional images |
|---|---|---|
| Colour | Grayscale | Colourful or grayscale |
| Dynamic range | Approximately $-3\times10^4$ to $3\times10^4$ | 0 to 255 |
| Colour depth(s) | 1 | 3 |
| Frequency content | Typically, 5 to 100 Hz | Wider band |

Table 2 Computational times of the customized U-Net and industry standard algorithm.

| Method | The customized U-Net | Industry standard algorithm |
|---|---|---|
| Computational time (per shot gather) | Extra 7.5 hours for training | Around 4s |
| | 0.02s | |